\documentclass[conference]{IEEEtran}
\IEEEoverridecommandlockouts
\usepackage{cite}
\usepackage{amsmath,amssymb,amsfonts}
\usepackage{algorithmic}
\usepackage{graphicx}
\usepackage[braket, qm]{qcircuit}
\usepackage{textcomp}
\usepackage{xcolor}
\usepackage{balance}
\usepackage{tikz}
\usepackage{pgfplots}
\usepackage{subcaption}
\pgfplotsset{compat=newest}
\usepackage{pgfplotstable}
\usepackage{url}
\usepackage{tikzit}
\tikzstyle{X dot}=[fill={rgb,255: red,255; green,136; blue,136}, tikzit category=nodes, shape=circle, draw=black, minimum size=20pt, inner sep=1pt]
\tikzstyle{Z dot}=[fill={rgb,255: red,204; green,255; blue,204}, tikzit category=nodes, shape=circle, draw=black, minimum size=20pt, inner sep=1pt]

\tikzstyle{Z dot dashed}=[dashed, fill={rgb,255: red,204; green,255; blue,204}, fill opacity=0.4, tikzit category=nodes, shape=circle, draw=black, minimum size=20pt, inner sep=1pt]

\tikzstyle{X dot rounded rectangle}=[fill={rgb,255: red,255; green,136; blue,136}, tikzit category=nodes, draw=black, shape=rounded rectangle, minimum size=20pt, inner sep=1pt]
\tikzstyle{Z dot rounded rectangle} = [fill={rgb,255: red,204; green,255; blue,204}, tikzit category=nodes, draw=black, shape=rounded rectangle, minimum size=20pt, inner sep=1pt]
\tikzstyle{hadamard}=[fill={rgb,255: red,255; green,255; blue,0}, shape=rectangle, draw=black]

\tikzstyle{hadamard edge}=[-, dashed, color={rgb,255:red,0;green,119;blue,255}, dash pattern=on 2pt off 1pt, line width=0.8pt]
\tikzstyle{dashed edge}=[-, thick, dashed]
\tikzset{every picture/.style={line width=1.0pt}}
\usepackage{float}

\usepackage{listings}
\usepackage{xcolor}

\definecolor{lightgray}{rgb}{0.95,0.95,0.95}
\definecolor{darkgray}{rgb}{0.35,0.35,0.35}
\definecolor{keywordblue}{RGB}{33,74,135}
\definecolor{functionpurple}{RGB}{153,0,153}
\definecolor{stringred}{RGB}{196,26,22}
\definecolor{commentgray}{RGB}{92,99,106}
\definecolor{numberorange}{RGB}{160,90,0}

\lstdefinelanguage{cypher}{
  morekeywords={
    MATCH,RETURN,WHERE,CREATE,DELETE,DETACH,SET,MERGE,WITH,
    AS,ORDER,BY,SKIP,LIMIT,UNWIND,DISTINCT,ON,REMOVE,FOREACH,
    CASE,WHEN,THEN,ELSE,END,REDUCE,ALL,ANY,NONE,SINGLE,IN,AND,OR,NOT
  },
  sensitive=true,
  morecomment=[l]{//},
  morecomment=[s]{/*}{*/},
  morestring=[b]",
  morestring=[b]',
  keywordstyle=\color{keywordblue}\bfseries,
  commentstyle=\color{commentgray}\itshape,
  stringstyle=\color{stringred},
  alsoletter={:},
  literate=
    *{0}{{{\color{numberorange}0}}}{1}
     {1}{{{\color{numberorange}1}}}{1}
     {2}{{{\color{numberorange}2}}}{1}
     {3}{{{\color{numberorange}3}}}{1}
     {4}{{{\color{numberorange}4}}}{1}
     {5}{{{\color{numberorange}5}}}{1}
     {6}{{{\color{numberorange}6}}}{1}
     {7}{{{\color{numberorange}7}}}{1}
     {8}{{{\color{numberorange}8}}}{1}
     {9}{{{\color{numberorange}9}}}{1}
     {reduce}{{{\color{functionpurple}\bfseries reduce}}}{6}
     {none}{{{\color{functionpurple}\bfseries none}}}{4}
     {any}{{{\color{functionpurple}\bfseries any}}}{3}
     {single}{{{\color{functionpurple}\bfseries single}}}{6}
     {all}{{{\color{functionpurple}\bfseries all}}}{3},
}

\usepackage[colorinlistoftodos]{todonotes}

\usepackage[defaultlines=3,all]{nowidow}
\usepackage{xspace}
\newcommand{\sys}{\textsc{ZX-db}\xspace}

\usepackage{hyperref}

\begin{document}


\title{\sys: A Graph Database for Quantum Circuit Simplification and Rewriting via the \emph{ZX-Calculus}}

\author{\IEEEauthorblockN{Valter Uotila}
\IEEEauthorblockA{\textit{Aalto University} \\
\textit{University of Helsinki}\\
valter.uotila@aalto.fi}
\and
\IEEEauthorblockN{Cong Yu}
\IEEEauthorblockA{\textit{Aalto University} \\
cong.yu@aalto.fi}
\and
\IEEEauthorblockN{Bo Zhao}
\IEEEauthorblockA{\textit{Aalto University} \\
bo.zhao@aalto.fi}
}

\maketitle

\begin{abstract}

Quantum computing is an emerging computational paradigm with the potential to outperform classical computers in solving a variety of problems. To achieve this, quantum programs are typically represented as quantum circuits, which must be optimized and adapted for target hardware through \emph{quantum circuit compilation}. 
We introduce \emph{\textbf{\sys}}, a data-driven system that performs \emph{quantum circuit simplification and rewriting} inside a graph database using \emph{ZX-calculus}, a complete graphical formalism for quantum mechanics. \sys encodes ZX-calculus rewrite rules as standard \emph{openCypher} queries and executes them on an example graph database engine, \emph{Memgraph}, enabling efficient, database-native transformations of large-scale quantum circuits. 
\sys integrates correctness validation via tensor and graph equivalence checks and is evaluated against the state-of-the-art \emph{PyZX} framework. 
Experimental results show that \sys achieves up to an order-of-magnitude speedup for independent rewrites, while exposing pattern-matching bottlenecks in current graph database engines. By uniting quantum compilation and graph data management, \sys opens a new systems direction toward scalable, database-supported quantum computing pipelines. \looseness=-1
\end{abstract}


\section{Introduction}

Quantum technologies are a rapidly emerging set of future technologies. The field can be divided into three high-level categories: quantum computing, distributed quantum systems, and quantum sensing. Among these technologies, quantum computing has already been applied to multiple database and query optimization problems~\cite{DBLP:conf/vldb/KesarwaniH24,Kittelmann2024Card,Uotila2024QueryMetrics,Gruenwald2023Index,Barbosa2024QRLIT,Trummer2024Index,Kesarwani2024Index,Groppe2021TSGrover,Bittner2020IDEAS,Bittner2020OJCC,Franz2024HypeQCE,Winker2023QMLJOO,schoenberger:23:pvldb,schoenberger:23:qdsm,Schoenberger:2023:sigmod,Winker2023Tut,Calikyilmaz2023Opp,Uotila_paper1,Uotila_paper4,Uotila_paper5,Uotila_Sahri_Groppe_2026,Uotila_Lu_2023}. While quantum computing is a promising technology and might play a significant role in optimizing future database management systems, this work focuses how databases can support quantum computing pipelines. Future quantum computational systems will likely create new, demanding data-processing problems that require the development of novel data management technologies and systems.\looseness=-1

Research on integrating modern databases and data management systems into quantum computing pipelines is still in its early stages. Interestingly, the well-known IKKBZ algorithm~\cite{Neumann_Radke_2018,10.1145/1270.1498} for join order optimization provides a linear-time algorithm also for computing optimal tensor network contraction paths~\cite{Stoian_Milbradt_Mendl_2024}. Tensor networks are widely used in quantum circuit simulation. Qymera~\cite{10.1145/3722212.3725126} is a system that demonstrates how quantum circuit simulation can be performed in relational databases. The most relevant system regarding this work is Pandora~\cite{moflic2025ultralargescalecompilationmanipulationquantum,Moflic_2024}, which utilizes PostgreSQL~\cite{50912,10.5555/190956.190989} to store quantum circuits and then efficiently performs specific quantum circuit transpilation steps at scale. We already proposed the idea of using graph databases for quantum circuit transpilation in~\cite{10.1145/3736393.3736694}.

In this work, we focus on one of the most crucial layers between quantum hardware and software: quantum circuit compilation~\cite{10.1007/978-3-031-90200-0_9,Meijer_van_de_Griend}. Like classical software, quantum algorithms and software must be compiled for quantum hardware. Compilation is usually expressed in terms of quantum circuits, which consist of wires describing quantum logical bits and quantum logic operations called gates. A complete compilation process consists of multiple stages, which modify quantum circuits. These stages include collecting circuit statistics, such as the number of gates, simplifying circuits, routing them to the target quantum hardware, synthesizing them, rewriting circuits using the hardware's native gate set, optimizing the number and types of gates, and scheduling operations. The order of the compilation stages is not fixed, and results from previous stages affect those in subsequent phases. Compilation can also account for the hardware's errors. Moreover, some optimization steps in quantum compilation are NP-hard and difficult even in practice. The compilation becomes even more challenging when one also considers that there are numerous competing quantum hardware technologies, each with its own compilation pipeline. Currently, compilation is mainly targeted for so-called Noisy Intermediate-Scale Quantum (NISQ) devices~\cite{Preskill_2018,8382253}. In contrast, quantum algorithms will also be compiled for fault-tolerant devices, and this compilation process will likely differ from that for NISQ devices. While there will be differences, the theoretical basis of this work offers elements that will also be useful in the fault-tolerant era~\cite{deBeaudrap2020zxcalculusis}.\looseness=-1

Whereas classical compilation is a mature field with standardized abstraction layers that enable software development without worrying about how the code is compiled and optimized, compilation for the current quantum hardware is relatively far from that. In this work, we explore the potential of graph databases for supporting the quantum compilation process. Graphs often serve as natural abstractions that bridge quantum circuits and compilation algorithms, motivating a deeper exploration of using graph databases in the compilation process. Since in the future the number of qubits will increase in millions and the number of gate operations is likely hundreds of millions, the systems that optimize graphs of possibly billions of nodes have to be scalable. In this regard, graph databases appear as promising candidates. They also automatically offer features, such as persistence, recoverability, and multi-user support, that quantum computing ecosystems might require in the future.

We identify compilation steps that appear viable to implement in current graph databases, while also discussing features that current graph databases should have but lack in this application domain. One of the most extensive and promising formalisms connecting quantum circuits and graphs is ZX-calculus, which is a comprehensive graph- and diagram-based framework for reasoning about quantum circuits and quantum mechanics~\cite{coecke2015generalised, ContPhys, CD1, Coecke2007graphicalcalculus}. The ZX-calculus is an especially useful formalism since it is complete for quantum mechanics~\cite{Jeandel_Perdrix_Vilmart_2020}. The curated list of publications using ZX-calculus is~\cite{zxcalculus_publist}, which covers over 360 articles. The core of the calculus is centered around a set of graph rewrite rules~\cite{Wetering_2020}. We demonstrate the feasibility of the graph databases in this topic by translating the ZX-calculus graph rewrite rules into openCypher graph queries~\cite{10.1145/3183713.3190657}. In practice, this means that ZX-calculus-based quantum circuit simplification can be performed by executing graph queries. The contributions of this work are as follows.
\begin{itemize}
    \item We have comprehensively analyzed the landscape of quantum circuit compilation, selected a well-defined set of functionalities, and constructed a collection of openCypher queries that perform quantum circuit simplification within property graph databases.
    \item We have benchmarked the \sys implementation against the baseline system, PyZX~\cite{Kissinger_2020}. The implementation is open-source and available on GitHub~\cite{valterUo2025zxdb}.
    \item The results divide the queries and rewrite rules into two categories that perform differently on graph databases. One set of queries outperforms or matches the baseline system's performance. In contrast, the other set requires querying disjoint graph patterns, which is not yet efficiently supported by graph databases. 
\end{itemize}


\section{Background}

\subsection{Software perspective for circuit compilation}

We begin by analyzing the problem landscape, presenting an overview, and defining the terminology used in this paper. Then, we narrow the scope to a few well-defined compilation phases and describe them in detail.

\begin{figure}
    \centering
    \input{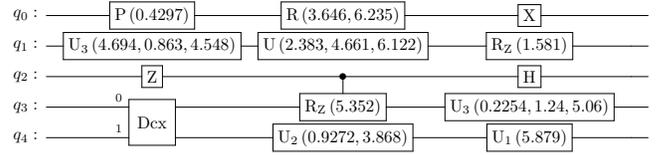}
    \caption{A random circuit generated with Qiskit}
    \label{fig:example_circuit}
\end{figure}

Due to space constraints, we aim to describe \sys in a way that requires a minimal background in quantum computing. Quantum computing is usually expressed in terms of quantum bits, i.e., qubits, that are unit vectors in a complex-valued Hilbert space. Considering the circuit example in Figure~\ref{fig:example_circuit}, the horizontal wires represent qubits. Quantum computation proceeds by manipulating qubits using quantum logical operations, known as gates. These gates are formally defined as unitary matrices. A circuit implements a quantum operation or an algorithm that can have multiple operationally equivalent representations as a circuit.

Since quantum computing is rapidly evolving and still lacks standards, the terminology for compilation varies slightly across communities and software. For example, Qiskit SDK implements transpilers~\cite{qiskit-transpiler-doc}, Pennylane executes transformations~\cite{pennylane-transforms-doc}, Cirq defines transformers~\cite{cirq-transform-doc} and Tket describes compilation~\cite{tket-comp-doc}. PyZX~\cite{PyZX-doc, Kissinger_2020}, which is crucial to this work, employs terms optimization and simplification. The quantum software landscape for compilers is generally vast. In addition to the previously mentioned packages, other systems that include a notable quantum circuit compiler are quizx~\cite{quizx}, Qiskit AI-powered Transpiler Service~\cite{qts-transpiler-doc}, BQSkit~\cite{BQSkit-passes-doc}, Quartz~\cite{Quartz-github}, QAT~\cite{QAT-github}, staq~\cite{staq-github}, ProjectQ~\cite{ProjectQ-github}, Quilc~\cite{quilc-github}, Unitary Compiler Collection~\cite{UCC-github}, Munich Quantum Toolkit~\cite{burgholzer2025MQTCore}, and Amazon Braket~\cite{AmazonBraket}.

\begin{figure}
    \centering
    \input{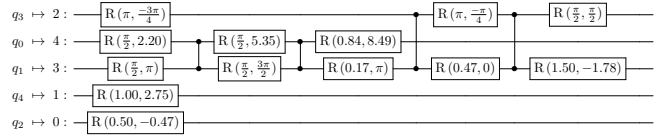}
    \caption{Compiled circuit in Fig.~\ref{fig:example_circuit}}
    \label{fig:transpiled_example}
\end{figure}

An example of a quantum circuit with random gate operations is presented in Fig.~\ref{fig:example_circuit}. The corresponding compiled circuit is given in Fig.~\ref{fig:transpiled_example}. The circuit is compiled for the IQM Adonis hardware using Qiskit, which is one of the most popular quantum software SDKs maintained by the Qiskit community and IBM. One can see that the compiled circuit contains only gates native to the IQM hardware. The so-called qubit mapping from the original circuit to the compiled circuit is presented on the left-hand side. For example, the original qubit $q_3$ is mapped to the physical qubit $2$ in the device. Qiskit also performs other optimization routines.

To better understand the functionalities supported by current quantum circuit compilation software, we have reviewed the core compilation functionalities of the main SDKs Qiskit, Pennylane, Cirq, Tket, and PyZX. Based on their documentations~\cite{qiskit-transpiler-doc, pennylane-transforms-doc, cirq-transform-doc, tket-comp-doc, PyZX-doc}, we classify the compilation functionalities as follows. \looseness=-1

\begin{itemize}
    \item \textbf{Circuit analysis}: Before the actual compilation process, it is often necessary to count the number of gates, their types, the circuit's depth, the circuit's width, etc. This information supports compilation and affects the selected compilation methods. 
    \item \textbf{Routing}: Routing refers to the process of constructing a mapping from the original qubits in the circuit to the physical qubits on the hardware. Since the hardware rarely implements the same connectivity between the qubits as the original circuit, it is often necessary to move the qubits, for example, by applying SWAP gates.
    \item \textbf{Decomposition}: By decomposition, we mean expanding complex gates into simpler gates without necessarily taking into account hardware constraints or optimization-related metrics. One of the most well-known decompositions is the decomposition for Toffoli-gate~\cite{Nielsen_Chuang_2010} in terms of single and two-qubit gates.
    \item \textbf{Optimization}: By optimization, we mean those compilation algorithms and steps where the focus is on rewriting circuits with respect to some performance metrics (e.g., fewer gates, shorter depth) at \textit{the circuit level} so that the goal is to minimize a specific metric.
    \item \textbf{Simplification}: Simplification steps focus on structural clarity or intermediate transformations without taking into account hardware constraints or other optimization constraints. Simplifications can be further classified into two categories: those made to the circuit itself and those to its metadata, such as renaming qubits, operations, registers, or removing visual barriers.
    \item \textbf{Rebase and squash}: These operations fix a gate set and rewrite the circuit accordingly. The common operation is to translate the circuit into the hardware's native gate set. The term "squash" is used when limiting the scope to local single or two-qubit gates.
    \item \textbf{Resynthesis}: Focus on rewriting circuits, possibly with respect to performance metrics and hardware constraints, using more abstract representations such as graphs, ZX-diagrams, or matrices. Often operates at a global level. This might be viewed as an upper class for optimization, simplification, decomposition, and rebasing.
    \item \textbf{Scheduling}: Some current quantum hardware and, in the future, all quantum hardware will support dynamic circuit capabilities, which allow classical control-flow and feedback based on mid-circuit measurement results. These operations require scheduling. The other class is that, surprisingly, quantum computers perform better if qubits do not have idle periods~\cite{Ezzell_Pokharel_Tewala_Quiroz_Lidar_2023}. This means one should perform gates without logical effect to suppress noise.\looseness=-1
\end{itemize}

\subsection{Problem definition and scope}

In this work, we focus on simplifications, particularly those implemented using ZX-calculus rules. While circuit analysis, circuit decomposition, and rebasing would also be viable steps to implement on graph databases, the simplification process is a more advanced and thus more motivating step to implement. We focus on using graph databases and graph queries and rely on existing simplification algorithms. Thus, the idea is not to implement better simplification algorithms but to express existing methods in a graph database-friendly format. To understand how ZX-calculus performs the simplifications and what kind of graph rewriting requirements it creates, we next briefly describe it.


\subsection{ZX-calculus}

\begin{figure}[H]
    \centering
    \resizebox{\columnwidth}{!}{
\begin{tikzpicture}
	\begin{pgfonlayer}{nodelayer}
		\node [style=Z dot] (0) at (0, 1) {$\alpha$};
		\node [style=X dot] (1) at (0, -2) {$\alpha$};
		\node [style=none] (2) at (-2, 2.25) {};
		\node [style=none] (3) at (-2, 1.5) {};
		\node [style=none] (4) at (-2, 0) {};
		\node [style=none] (5) at (-2, -0.75) {};
		\node [style=none] (6) at (-2, -1.5) {};
		\node [style=none] (7) at (2, 2.25) {};
		\node [style=none] (8) at (2, 1.5) {};
		\node [style=none] (9) at (2, 0) {};
		\node [style=none] (10) at (-2, -3) {};
		\node [style=none] (11) at (2, -0.75) {};
		\node [style=none] (12) at (2, -1.5) {};
		\node [style=none] (13) at (2, -3) {};
		\node [style=none] (14) at (2, -2) {$\vdots$};
		\node [style=none] (15) at (-2, -2) {$\vdots$};
		\node [style=none] (16) at (2, 1) {$\vdots$};
		\node [style=none] (17) at (-2, 1) {$\vdots$};
		\node [style=none] (18) at (2.75, 1) {$=$};
		\node [style=none] (20) at (5.85, 1) {$|0 \cdots 0 \rangle\langle 0 \cdots 0| + e^{ i \alpha} | 1 \cdots 1 \rangle \langle1 \cdots 1|$};
		\node [style=none] (21) at (2.75, -2) {$=$};
		\node [style=none] (22) at (6.5, -2) {$|+ \cdots + \rangle\langle + \cdots +| + e^{ i \alpha} | - \cdots - \rangle \langle- \cdots -|$};
	\end{pgfonlayer}
	\begin{pgfonlayer}{edgelayer}
		\draw (0) to (7.center);
		\draw (0) to (8.center);
		\draw (0) to (9.center);
		\draw (2.center) to (0);
		\draw (3.center) to (0);
		\draw (4.center) to (0);
		\draw (1) to (5.center);
		\draw (1) to (11.center);
		\draw (1) to (12.center);
		\draw (1) to (6.center);
		\draw (1) to (10.center);
		\draw (1) to (13.center);
	\end{pgfonlayer}
\end{tikzpicture}
}
    \caption{Generators for $Z$- and $X$-spiders with phase $\alpha$}
    \label{fig:spider_definitions}
\end{figure}
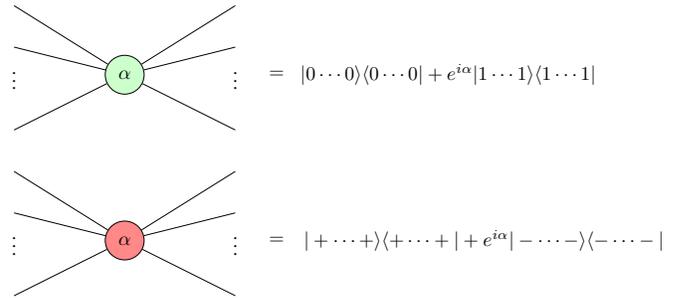

There is extensive previous research on ZX-calculus, which covers how the ZX-diagrams can be used to express and rewrite quantum circuits~\cite{Wetering_2020}. Given a quantum circuit with arbitrary operations, one can always turn it into the equivalent ZX-diagram, i.e., a graph. Every ZX-diagram is generated by a small set of operations, with the most important being $Z$- and $X$-spiders, denoted by green and red nodes, respectively. The generators for these spiders are given in Fig.~\ref{fig:spider_definitions}. The spiders also have a phase value, e.g., $\alpha$ in Fig.~\ref{fig:spider_definitions}, which describes a rotation around the $Z$-axis or $X$-axis, respectively. Generally, a ZX-diagram is a particular tensor network, and the correctness of the transformations in \sys will be partly based on tensor networks.

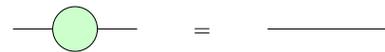
\begin{figure}[H]
    \centering
    \resizebox{0.6\columnwidth}{!}{
\begin{tikzpicture}
    \begin{pgfonlayer}{nodelayer}
        \node [style=Z dot] (0) at (-9.00, 12.25) {};
        \node [style=none] (1) at (-10.00, 12.25) {};
        \node [style=none] (2) at (-8.00, 12.25) {};
\node [style=none] (4) at (-7.00, 12.2) {$=$};
        \node [style=none] (5) at (-6.00, 12.25) {};
        \node [style=none] (6) at (-4.00, 12.25) {};
    \end{pgfonlayer}
    \begin{pgfonlayer}{edgelayer}
        \draw (0) to (1);
        \draw (0) to (2);
        \draw (5) to (6);
    \end{pgfonlayer}
\end{tikzpicture}
}
    \caption{Identity removal}
    \label{fig:remove_identity}
\end{figure}

Next, we focus on the rewrite rules that \sys implements with property graph queries. We start with simple rules and proceed to more complex pattern matching and rewriting. The first rule is a so-called identity removal rule presented in Fig.~\ref{fig:remove_identity}. This rule states that a green spider with a phase value of $0$ and degree $2$ can be removed.

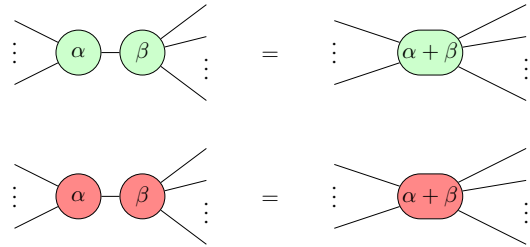
\begin{figure}[H]
    \centering
    \resizebox{0.8\columnwidth}{!}{
\begin{tikzpicture}
	\begin{pgfonlayer}{nodelayer}
		\node [style=none] (0) at (-4, 1.75) {};
		\node [style=Z dot] (1) at (-3, 1.25) {$\alpha$};
		\node [style=Z dot] (2) at (-2, 1.25) {$\beta$};
		\node [style=none] (3) at (-1, 1.5) {};
		\node [style=none] (4) at (-1, 0.5) {};
		\node [style=none] (5) at (-4, 0.75) {};
		\node [style=none] (6) at (-1, 2) {};
		\node [style=none] (7) at (-1, -0.25) {};
		\node [style=none] (8) at (-4, -0.5) {};
		\node [style=none] (9) at (-1, -0.75) {};
		\node [style=none] (10) at (-4, -1.5) {};
		\node [style=none] (11) at (-1, -1.75) {};
		\node [style=X dot] (12) at (-3, -1) {$\alpha$};
		\node [style=X dot] (13) at (-2, -1) {$\beta$};
		\node [style=none] (17) at (0, -1.05) {$=$};
		\node [style=none] (17) at (0, 1.2) {$=$};
		\node [style=none] (18) at (1, 1.75) {};
		\node [style=Z dot rounded rectangle] (19) at (2.5, 1.25) {$\alpha + \beta$};
		\node [style=none] (20) at (4, 1.5) {};
		\node [style=none] (21) at (4, 0.5) {};
		\node [style=none] (22) at (1, 0.75) {};
		\node [style=none] (23) at (4, 2) {};
		\node [style=none] (24) at (4, -0.25) {};
		\node [style=none] (25) at (1, -0.5) {};
		\node [style=none] (26) at (4, -0.75) {};
		\node [style=none] (27) at (1, -1.5) {};
		\node [style=none] (28) at (4, -1.75) {};
		\node [style=X dot rounded rectangle] (29) at (2.5, -1) {$\alpha + \beta$};
		\node [style=none] (30) at (4, -1.15) {$\vdots$};
		\node [style=none] (31) at (4, 1.1) {$\vdots$};
		\node [style=none] (32) at (1, 1.35) {$\vdots$};
		\node [style=none] (33) at (1, -0.9) {$\vdots$};
		\node [style=none] (34) at (-4, -0.9) {$\vdots$};
		\node [style=none] (35) at (-4, 1.35) {$\vdots$};
		\node [style=none] (36) at (-1, -1.15) {$\vdots$};
		\node [style=none] (37) at (-1, 1.1) {$\vdots$};
	\end{pgfonlayer}
	\begin{pgfonlayer}{edgelayer}
		\draw (0.center) to (1);
		\draw (1) to (2);
		\draw (1) to (5.center);
		\draw (2) to (3.center);
		\draw (2) to (6.center);
		\draw (2) to (4.center);
		\draw (7.center) to (13);
		\draw (8.center) to (12);
		\draw (9.center) to (13);
		\draw (10.center) to (12);
		\draw (11.center) to (13);
		\draw (12) to (13);
		\draw (18.center) to (19);
		\draw (19) to (20.center);
		\draw (19) to (21.center);
		\draw (19) to (22.center);
		\draw (19) to (23.center);
		\draw (24.center) to (29);
		\draw (25.center) to (29);
		\draw (26.center) to (29);
		\draw (27.center) to (29);
		\draw (28.center) to (29);
	\end{pgfonlayer}
\end{tikzpicture}
}
    \caption{Spiders with the same color can be fused and their phases add}
    \label{fig:spider_fusion}
\end{figure}

The following rule states that connected spiders with the same color fuse together, and their phases add. This rule is demonstrated in Fig.~\ref{fig:spider_fusion}.

\begin{figure}[H]
    \centering
    \resizebox{0.8\columnwidth}{!}{
\begin{tikzpicture}
	\begin{pgfonlayer}{nodelayer}
		\node [style=Z dot] (0) at (1.75, -0.25) {};
		\node [style=X dot] (1) at (3, -0.25) {};
		\node [style=none] (2) at (4.25, 1) {};
		\node [style=none] (3) at (4.25, 0.25) {};
		\node [style=none] (4) at (4.25, -1.25) {};
		\node [style=none] (5) at (0.75, 0.75) {};
		\node [style=none] (6) at (0.75, -1) {};
		\node [style=none] (8) at (0, -0.3) {$=$};
		\node [style=none] (9) at (-1, 1) {};
		\node [style=none] (10) at (-1, 0) {};
		\node [style=none] (11) at (-1, -1.75) {};
		\node [style=none] (12) at (-5, 1) {};
		\node [style=none] (13) at (-5, -1.25) {};
		\node [style=Z dot] (14) at (-2, 1) {};
		\node [style=Z dot] (15) at (-2, 0) {};
		\node [style=Z dot] (16) at (-2, -1.75) {};
		\node [style=X dot] (17) at (-4.05, 0.55) {};
		\node [style=X dot] (18) at (-4, -0.75) {};
		\node [style=none] (19) at (0.75, 0) {$\vdots$};
		\node [style=none] (20) at (-5, 0) {$\vdots$};
		\node [style=none] (21) at (4, -0.25) {$\vdots$};
		\node [style=none] (23) at (-2, -0.75) {$\vdots$};
	\end{pgfonlayer}
	\begin{pgfonlayer}{edgelayer}
		\draw (0) to (1);
		\draw (0) to (5.center);
		\draw (0) to (6.center);
		\draw (1) to (2.center);
		\draw (1) to (3.center);
		\draw (1) to (4.center);
		\draw (9.center) to (14);
		\draw (10.center) to (15);
		\draw (11.center) to (16);
		\draw (12.center) to (17);
		\draw (13.center) to (18);
		\draw (14) to (17);
		\draw (14) to (18);
		\draw (15) to (17);
		\draw (15) to (18);
		\draw (16) to (17);
		\draw (16) to (18);
	\end{pgfonlayer}
\end{tikzpicture}
}
    \caption{An example of a bialgebra rule}
    \label{fig:bialgebra}
\end{figure}
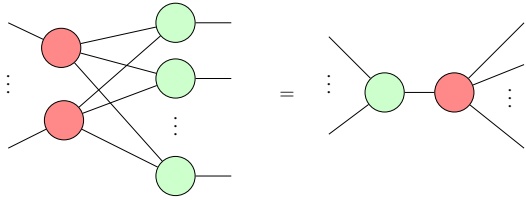

Fig.~\ref{fig:bialgebra} demonstrates the so-called bialgebra rule, which states that a green-red pair of spiders, connected with a single edge, can be turned into a bipartite graph so that we possibly introduce more green and red spiders and all-to-all connectivity between them. Fig.~\ref{fig:bialgebra} illustrates the rule that aligns with the direction of the bipartite graph and then simplifies it, removing many nodes and edges.

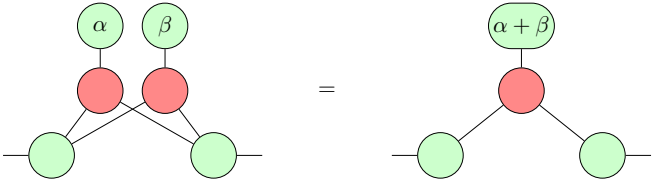
\begin{figure}[H]
    \centering
    \resizebox{\columnwidth}{!}{
\begin{tikzpicture}
	\begin{pgfonlayer}{nodelayer}
		\node [style=Z dot] (0) at (-4.25, -1) {};
		\node [style=Z dot] (1) at (-1.75, -1) {};
		\node [style=Z dot] (2) at (-3.5, 1) {$\alpha$};
		\node [style=Z dot] (3) at (-2.5, 1) {$\beta$};
		\node [style=X dot] (4) at (-3.5, 0) {};
		\node [style=X dot] (5) at (-2.5, 0) {};
		\node [style=none] (6) at (-1, -1) {};
		\node [style=none] (7) at (-5, -1) {};
		\node [style=none] (9) at (0, 0) {$=$};
		\node [style=Z dot] (10) at (1.75, -1) {};
		\node [style=Z dot] (11) at (4.25, -1) {};
		\node [style=Z dot rounded rectangle] (12) at (3, 1) {$\alpha + \beta$};
		\node [style=X dot] (13) at (3, 0) {};
		\node [style=none] (14) at (5, -1) {};
		\node [style=none] (15) at (1, -1) {};
	\end{pgfonlayer}
	\begin{pgfonlayer}{edgelayer}
		\draw (0) to (4);
		\draw (0) to (5);
		\draw (0) to (7.center);
		\draw (1) to (4);
		\draw (1) to (5);
		\draw (1) to (6.center);
		\draw (2) to (4);
		\draw (3) to (5);
		\draw (10) to (13);
		\draw (10) to (15.center);
		\draw (11) to (13);
		\draw (11) to (14.center);
		\draw (12) to (13);
	\end{pgfonlayer}
\end{tikzpicture}

}
    \caption{Gadget fusion example}
    \label{fig:gadget_fusion}
\end{figure}

The following rule, known as the gadget fusion rule, states that two red spiders sharing a pair of green spiders, each connected to an otherwise isolated green spider, can be fused together. In this process, the phases of the green spiders are added. Fig.~\ref{fig:gadget_fusion} demonstrates this. The rule naturally generalizes to any sequence of this pattern since it can be applied pairwise. 

All the previous rules have contained so-called simple edges, which we have denoted with a black wire. The following rules will introduce a new edge, called a Hadamard edge, which we denote with a blue dashed line. These edges are based on one of the most used quantum logical gates, the Hadamard gate, which is used, for example, to introduce superposition. Since it is a relatively common operation, it has its own edge, which simplifies the notation. The exact definition is given in Fig.~\ref{fig:hadamard_edge}.

\begin{figure}[H]
    \centering
    \resizebox{\columnwidth}{!}{
\begin{tikzpicture}
	\begin{pgfonlayer}{nodelayer}
		\node [style=none] (0) at (-1.25, 0) {};
		\node [style=none] (1) at (1.5, 0) {};
		\node [style=hadamard] (2) at (0, 0) {};
		\node [style=none] (3) at (-2, 0) {$=$};
		\node [style=none] (4) at (2, 0) {$=$};
		\node [style=none] (5) at (-4.25, 0) {};
		\node [style=none] (6) at (-2.75, 0) {};
		\node [style=none] (7) at (3.75, 0) {$\frac{1}{\sqrt{2}}
\begin{pmatrix}
1 & 1 \\
1 & -1
\end{pmatrix}$};
	\end{pgfonlayer}
	\begin{pgfonlayer}{edgelayer}
		\draw (0.center) to (2);
		\draw (2) to (1.center);
		\draw [style=hadamard edge] (5.center) to (6.center);
	\end{pgfonlayer}
\end{tikzpicture}
}
    \caption{Definition of Hadamard edge as a yellow Hadamard gate which corresponds to the Hadamard matrix}
    \label{fig:hadamard_edge}
\end{figure}

\begin{figure}[H]
    \centering
    \resizebox{\columnwidth}{!}{

\begin{tikzpicture}
	\begin{pgfonlayer}{nodelayer}
		\node [style=none] (0) at (-6.25, -1) {};
		\node [style=Z dot] (1) at (-5, 1.5) {$\alpha_1$};
		\node [style=Z dot] (2) at (-5.5, 0.25) {$\alpha_2$};
		\node [style=Z dot] (3) at (-2.5, 1.5) {$\alpha_6$};
		\node [style=Z dot] (4) at (-2, 0.25) {$\alpha_5$};
		\node [style=none] (5) at (-6.25, 0.25) {};
		\node [style=Z dot] (6) at (-3.75, 0.25) {$\pm \frac{\pi}{2}$};
		\node [style=Z dot] (7) at (-5, -1) {$\alpha_3$};
		\node [style=Z dot] (8) at (-2.5, -1) {$\alpha_4$};
		\node [style=none] (9) at (-6.25, 1.5) {};
		\node [style=none] (10) at (-1, 1.5) {};
		\node [style=none] (11) at (-1, 0.25) {};
		\node [style=none] (12) at (-1, -1) {};
		\node [style=none] (14) at (0, 0) {$=$};
		\node [style=none] (15) at (1, -1) {};
		\node [style=Z dot rounded rectangle] (16) at (2.25, 1.5) {$\alpha_1 \mp \frac{\pi}{2}$};
		\node [style=Z dot rounded rectangle] (17) at (1.5, 0.25) {$\alpha_2 \mp \frac{\pi}{2}$};
		\node [style=Z dot rounded rectangle] (18) at (4.75, 1.5) {$\alpha_6 \mp \frac{\pi}{2}$};
		\node [style=Z dot rounded rectangle] (19) at (5.5, 0.25) {$\alpha_5 \mp \frac{\pi}{2}$};
		\node [style=none] (20) at (0.5, 0.25) {};
		\node [style=Z dot rounded rectangle] (21) at (2.25, -1) {$\alpha_3 \mp \frac{\pi}{2}$};
		\node [style=Z dot rounded rectangle] (22) at (4.75, -1) {$\alpha_4 \mp \frac{\pi}{2}$};
		\node [style=none] (23) at (1, 1.5) {};
		\node [style=none] (24) at (6.25, 1.5) {};
		\node [style=none] (25) at (6.75, 0.25) {};
		\node [style=none] (26) at (6.25, -1) {};
	\end{pgfonlayer}
	\begin{pgfonlayer}{edgelayer}
		\draw (0.center) to (7);
		\draw [style=hadamard edge] (1) to (6);
		\draw (1) to (9.center);
		\draw [style=hadamard edge] (2) to (6);
		\draw (2) to (5.center);
		\draw [style=hadamard edge] (3) to (6);
		\draw (3) to (10.center);
		\draw [style=hadamard edge] (4) to (6);
		\draw [style=hadamard edge] (4) to (7);
		\draw (4) to (11.center);
		\draw [style=hadamard edge] (6) to (7);
		\draw [style=hadamard edge, in=135, out=-45] (6) to (8);
		\draw [style=hadamard edge] (7) to (8);
		\draw (8) to (12.center);
		\draw (15.center) to (21);
		\draw (16) to (23.center);
		\draw [style=hadamard edge] (16) to (17);
		\draw [style=hadamard edge] (16) to (18);
		\draw [style=hadamard edge] (16) to (19);
		\draw [style=hadamard edge] (16) to (21);
		\draw [style=hadamard edge] (16) to (22);
		\draw (17) to (20.center);
		\draw [style=hadamard edge] (17) to (18);
		\draw [style=hadamard edge] (17) to (19);
		\draw [style=hadamard edge] (17) to (21);
		\draw [style=hadamard edge] (17) to (22);
		\draw (18) to (24.center);
		\draw [style=hadamard edge] (18) to (19);
		\draw [style=hadamard edge] (18) to (21);
		\draw [style=hadamard edge] (18) to (22);
		\draw (19) to (25.center);
		\draw [style=hadamard edge] (19) to (22);
		\draw (22) to (26.center);
	\end{pgfonlayer}
\end{tikzpicture}

}
    \caption{Example of local complementation rule}
    \label{fig:local_complementation_rule}
\end{figure}
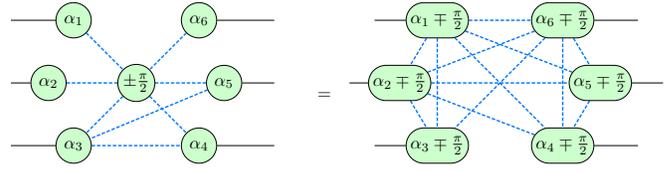

The first rule, expressed in terms of Hadamard edges, is known as the local complementation rule. We can construct the complement graph for the Hadamard edge–connected nodes linked to the green spider with a phase of $\pm \pi/2$, and then remove that $\pm \pi/2$ node. This operation corresponds to adding or removing a $\pi/2$ phase from its neighboring nodes. Fig.~\ref{fig:local_complementation_rule} demonstrates the rule for six neighboring nodes, but the rule naturally applies to any number of nodes.

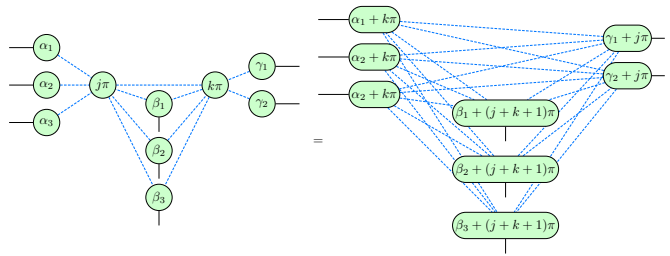
\begin{figure}[H]
    \centering
    \resizebox{\columnwidth}{!}{
\begin{tikzpicture}
	\begin{pgfonlayer}{nodelayer}
		\node [style=none] (0) at (-3.25, 0) {};
		\node [style=Z dot] (1) at (-3.25, -0.5) {$\beta_2$};
		\node [style=Z dot] (2) at (-0.5, 0.75) {$\gamma_2$};
		\node [style=none] (3) at (-7.25, 1.25) {};
		\node [style=none] (4) at (-7.25, 0.25) {};
		\node [style=Z dot] (5) at (-6.25, 0.25) {$\alpha_3$};
		\node [style=none] (6) at (0.5, 1.75) {};
		\node [style=none] (7) at (-3.25, -2.5) {};
		\node [style=Z dot] (8) at (-6.25, 2.25) {$\alpha_1$};
		\node [style=Z dot] (9) at (-6.25, 1.25) {$\alpha_2$};
		\node [style=none] (10) at (-3.25, -1.25) {};
		\node [style=Z dot] (11) at (-0.5, 1.75) {$\gamma_1$};
		\node [style=Z dot] (12) at (-1.75, 1.25) {$k\pi$};
		\node [style=Z dot] (13) at (-3.25, 0.75) {$\beta_1$};
		\node [style=Z dot] (14) at (-4.75, 1.25) {$j\pi$};
		\node [style=none] (15) at (-7.25, 2.25) {};
		\node [style=Z dot] (16) at (-3.25, -1.75) {$\beta_3$};
		\node [style=none] (17) at (0.5, 0.75) {};
		\node [style=none] (18) at (6, -0.25) {};
		\node [style=Z dot rounded rectangle] (19) at (6, -1) {$\beta_2 + (j + k + 1)\pi$};
		\node [style=Z dot rounded rectangle] (20) at (9.25, 1.5) {$\gamma_2 + j\pi$};
		\node [style=none] (21) at (1, 2) {};
		\node [style=none] (22) at (1, 1) {};
		\node [style=Z dot rounded rectangle] (23) at (2.5, 1) {$\alpha_2 + k\pi$};
		\node [style=none] (24) at (10.25, 2.5) {};
		\node [style=none] (25) at (6, -3.25) {};
		\node [style=Z dot rounded rectangle] (26) at (2.5, 3) {$\alpha_1 + k\pi$};
		\node [style=Z dot rounded rectangle] (27) at (2.5, 2) {$\alpha_2 + k\pi$};
		\node [style=none] (28) at (6, -1.75) {};
		\node [style=Z dot rounded rectangle] (29) at (9.25, 2.5) {$\gamma_1 + j\pi$};
		\node [style=Z dot rounded rectangle] (30) at (6, 0.5) {$\beta_1 + (j + k + 1)\pi$};
		\node [style=none] (31) at (1, 3) {};
		\node [style=Z dot rounded rectangle] (32) at (6, -2.5) {$\beta_3 + (j + k + 1)\pi$};
		\node [style=none] (33) at (10.25, 1.5) {};
		\node [style=none] (34) at (1, -0.25) {$=$};
	\end{pgfonlayer}
	\begin{pgfonlayer}{edgelayer}
		\draw (0.center) to (13);
		\draw (1) to (10.center);
		\draw [style=hadamard edge] (1) to (12);
		\draw [style=hadamard edge] (1) to (14);
		\draw (2) to (17.center);
		\draw [style=hadamard edge] (2) to (12);
		\draw (3.center) to (9);
		\draw (4.center) to (5);
		\draw [style=hadamard edge] (5) to (14);
		\draw (6.center) to (11);
		\draw (7.center) to (16);
		\draw (8) to (15.center);
		\draw [style=hadamard edge] (8) to (14);
		\draw [style=hadamard edge] (9) to (14);
		\draw [style=hadamard edge] (11) to (12);
		\draw [style=hadamard edge] (12) to (13);
		\draw [style=hadamard edge] (12) to (16);
		\draw [style=hadamard edge] (12) to (14);
		\draw [style=hadamard edge] (13) to (14);
		\draw [style=hadamard edge] (14) to (16);
		\draw (18.center) to (30);
		\draw (19) to (28.center);
		\draw [style=hadamard edge] (19) to (26);
		\draw [style=hadamard edge] (19) to (27);
		\draw [style=hadamard edge] (19) to (23);
		\draw [style=hadamard edge] (19) to (20);
		\draw [style=hadamard edge] (19) to (29);
		\draw (20) to (33.center);
		\draw [style=hadamard edge] (20) to (26);
		\draw [style=hadamard edge] (20) to (27);
		\draw [style=hadamard edge] (20) to (23);
		\draw [style=hadamard edge] (20) to (32);
		\draw [style=hadamard edge] (20) to (30);
		\draw (21.center) to (27);
		\draw (22.center) to (23);
		\draw [style=hadamard edge] (23) to (29);
		\draw [style=hadamard edge] (23) to (32);
		\draw [style=hadamard edge] (23) to (30);
		\draw (24.center) to (29);
		\draw (25.center) to (32);
		\draw (26) to (31.center);
		\draw [style=hadamard edge] (26) to (29);
		\draw [style=hadamard edge] (26) to (32);
		\draw [style=hadamard edge] (26) to (30);
		\draw [style=hadamard edge] (27) to (29);
		\draw [style=hadamard edge] (27) to (32);
		\draw [style=hadamard edge] (27) to (30);
		\draw [style=hadamard edge] (29) to (32);
		\draw [style=hadamard edge] (29) to (30);
	\end{pgfonlayer}
\end{tikzpicture}

}
    \caption{Example of the pivot rule}
    \label{fig:pivot_rule}
\end{figure}

Finally, we present a collection of so-called pivot-rules. To keep the graphs readable, we demonstrate the rules for a fixed number of nodes, but the rules automatically apply to graphs with varying node counts. The nodes whose number can vary are marked with variables. The first rule in the category of pivot rules is the pivot rule, which works as a basis for the other pivot rules. The rule is demonstrated in Fig.~\ref{fig:pivot_rule}. The identification of the pattern begins by finding two pivot nodes, which are green spiders with a phase that is an integer multiple of $\pi$. The pivot nodes are connected with a so-called pivot edge. The neighbors' set of the pivot nodes can be divided into three categories: those neighbors that are connected to the first pivot node (marked with variable $\alpha$ in Fig.~\ref{fig:pivot_rule}), those that are connected to the second pivot node (marked with variable $\gamma$ in Fig.~\ref{fig:pivot_rule}), and those that are linked to both (marked with variable $\beta$ in Fig.~\ref{fig:pivot_rule}). This type of pattern can be rewritten as a local complement graph, removing the pivot nodes and the pivot edge and adding certain phases to the neighboring nodes. Fig.~\ref{fig:pivot_rule} clarifies this process.

\begin{figure}[H]
    \centering
    \resizebox{\columnwidth}{!}{
\begin{tikzpicture}
	\begin{pgfonlayer}{nodelayer}
		\node [style=Z dot] (0) at (-2.5, 4.5) {$\varphi$};
		\node [style=Z dot] (4) at (-7, 5.25) {$\alpha_1$};
		\node [style=Z dot] (6) at (-4, 3.5) {$\beta_1$};
		\node [style=Z dot] (10) at (-5.5, 4.5) {$k\pi$};
		\node [style=Z dot] (12) at (-1, 4.75) {$\gamma_1$};
		\node [style=none] (14) at (0, 4.75) {};
		\node [style=none] (15) at (-8, 5.25) {};
		\node [style=none] (17) at (-4, 2.75) {};
		\node [style=none] (18) at (-4, 1.5) {};
		\node [style=Z dot] (19) at (-4, 2.25) {$\beta_2$};
		\node [style=none] (20) at (-4, 0) {};
		\node [style=Z dot] (21) at (-4, 0.75) {$\beta_3$};
		\node [style=Z dot] (22) at (-7, 4.5) {$\alpha_2$};
		\node [style=none] (23) at (-8, 4.5) {};
		\node [style=Z dot] (24) at (-1, 4) {$\gamma_2$};
		\node [style=none] (25) at (0, 4) {};
		\node [style=Z dot] (26) at (-7, 3.75) {$\alpha_3$};
		\node [style=none] (27) at (-8, 3.75) {};
		\node [style=Z dot rounded rectangle] (28) at (5, 3) {$\beta_1 + (k + 1)\pi$};
		\node [style=Z dot] (29) at (8.5, 5.75) {$\varphi$};
		\node [style=none] (30) at (5, -1) {};
		\node [style=Z dot rounded rectangle] (31) at (5, -0.25) {$\beta_3 + (k + 1)\pi$};
		\node [style=none] (32) at (0.5, 4) {};
		\node [style=Z dot] (33) at (1.5, 3.25) {$\alpha_3$};
		\node [style=Z dot rounded rectangle] (34) at (5, 1.5) {$\beta_2 + (k + 1)\pi$};
		\node [style=none] (35) at (9.5, 4) {};
		\node [style=none] (36) at (0.5, 3.25) {};
		\node [style=none] (37) at (5, 0.75) {};
		\node [style=Z dot] (38) at (1.5, 4.75) {$\alpha_1$};
		\node [style=none] (39) at (0.5, 4.75) {};
		\node [style=Z dot] (40) at (1.5, 4) {$\alpha_2$};
		\node [style=none] (41) at (5, 2.25) {};
		\node [style=none] (42) at (9.5, 4.75) {};
		\node [style=Z dot rounded rectangle] (43) at (8.5, 4.75) {$\gamma_1 + k\pi$};
		\node [style=Z dot rounded rectangle] (44) at (7, 5.75) {$k\pi$};
		\node [style=Z dot rounded rectangle] (45) at (8.5, 4) {$\gamma_2 + k\pi$};
		\node [style=none] (46) at (0, 2.25) {$=$};
	\end{pgfonlayer}
	\begin{pgfonlayer}{edgelayer}
		\draw [style=hadamard edge] (0) to (6);
		\draw [style=hadamard edge] (0) to (10);
		\draw [style=hadamard edge] (0) to (12);
		\draw [style=hadamard edge] (0) to (19);
		\draw [style=hadamard edge] (0) to (21);
		\draw [style=hadamard edge] (0) to (24);
		\draw (4) to (15.center);
		\draw [style=hadamard edge] (4) to (10);
		\draw [style=hadamard edge] (6) to (10);
		\draw (6) to (17.center);
		\draw [style=hadamard edge] (10) to (19);
		\draw [style=hadamard edge] (10) to (21);
		\draw [style=hadamard edge] (10) to (22);
		\draw [style=hadamard edge] (10) to (26);
		\draw (12) to (14.center);
		\draw (18.center) to (19);
		\draw (20.center) to (21);
		\draw (22) to (23.center);
		\draw (24) to (25.center);
		\draw (26) to (27.center);
		\draw [style=hadamard edge] (28) to (44);
		\draw (28) to (41.center);
		\draw [style=hadamard edge] (28) to (45);
		\draw [style=hadamard edge] (28) to (43);
		\draw [style=hadamard edge] (28) to (33);
		\draw [style=hadamard edge] (28) to (40);
		\draw [style=hadamard edge] (28) to (38);
		\draw [style=hadamard edge] (29) to (44);
		\draw (30.center) to (31);
		\draw [style=hadamard edge] (31) to (45);
		\draw [style=hadamard edge] (31) to (33);
		\draw [style=hadamard edge] (31) to (40);
		\draw [style=hadamard edge] (31) to (38);
		\draw [style=hadamard edge] (31) to (43);
		\draw [style=hadamard edge] (31) to (44);
		\draw (32.center) to (40);
		\draw (33) to (36.center);
		\draw [style=hadamard edge] (33) to (34);
		\draw [style=hadamard edge] (33) to (43);
		\draw [style=hadamard edge] (33) to (44);
		\draw [style=hadamard edge] (33) to (45);
		\draw [style=hadamard edge] (34) to (45);
		\draw [style=hadamard edge] (34) to (40);
		\draw (34) to (37.center);
		\draw [style=hadamard edge] (34) to (38);
		\draw [style=hadamard edge] (34) to (43);
		\draw [style=hadamard edge] (34) to (44);
		\draw (35.center) to (45);
		\draw (38) to (39.center);
		\draw [style=hadamard edge] (38) to (44);
		\draw [style=hadamard edge] (38) to (45);
		\draw [style=hadamard edge] (38) to (43);
		\draw [style=hadamard edge] (40) to (45);
		\draw [style=hadamard edge] (40) to (43);
		\draw [style=hadamard edge] (40) to (44);
		\draw (42.center) to (43);
	\end{pgfonlayer}
\end{tikzpicture}

}
    \caption{Example of the pivot gadget rule}
    \label{fig:pivot_gadget_rule}
\end{figure}
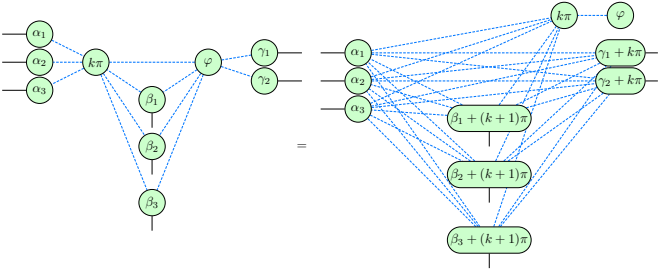

The second pivot-type rule is called a pivot gadget rule. This rule is similar to the pivot rule except that in this case, one of the pivot nodes does not need to have a phase that is an integer multiple of $\pi$. The idea for tackling this case is that the arbitrary phase value $\phi$ can be moved ''outside'' of the pivot pattern, and then the pivot rule can be applied. Fig.~\ref{fig:pivot_gadget_rule} demonstrates this rule.

\begin{figure}[H]
    \centering
    \resizebox{\columnwidth}{!}{
\begin{tikzpicture}
	\begin{pgfonlayer}{nodelayer}
		\node [style=Z dot] (0) at (-2.75, 2.5) {$\varphi$};
		\node [style=Z dot] (1) at (-6.5, 3.25) {$\alpha_1$};
		\node [style=Z dot] (2) at (-4, 1.5) {$\beta_1$};
		\node [style=Z dot] (3) at (-5.25, 2.5) {$k\pi$};
		\node [style=Z dot] (4) at (-1.5, 3) {$\gamma_1$};
		\node [style=none] (5) at (-0.5, 2.75) {};
		\node [style=none] (6) at (-7.5, 3.25) {};
		\node [style=none] (7) at (-5, 0.75) {};
		\node [style=none] (8) at (-4.75, -0.5) {};
		\node [style=Z dot] (9) at (-4, 0.5) {$\beta_2$};
		\node [style=Z dot] (12) at (-6.5, 2.5) {$\alpha_2$};
		\node [style=none] (13) at (-7.5, 2.75) {};
		\node [style=Z dot] (14) at (-1.5, 1.75) {$\gamma_2$};
		\node [style=none] (15) at (-0.5, 2) {};
		\node [style=Z dot] (16) at (-6.5, 1.75) {$\alpha_3$};
		\node [style=none] (17) at (-7.5, 1.75) {};
		\node [style=none] (18) at (-2.75, 3.75) {};
		\node [style=none] (19) at (-3, 0.5) {};
		\node [style=none] (20) at (-3.25, -0.5) {};
		\node [style=none] (22) at (-7.5, 3.75) {};
		\node [style=none] (23) at (-7.5, 1.25) {};
		\node [style=none] (24) at (-7.5, 2.25) {};
		\node [style=none] (25) at (-0.5, 1.5) {};
		\node [style=none] (26) at (-0.5, 3.25) {};
		\node [style=Z dot] (27) at (6.25, 4.25) {$k\pi$};
		\node [style=none] (28) at (8.75, 1.75) {};
		\node [style=none] (29) at (0.5, 3.5) {};
		\node [style=none] (31) at (0.5, 1.5) {};
		\node [style=Z dot rounded rectangle] (32) at (7.75, 2) {$\gamma_2 + k\pi$};
		\node [style=none] (33) at (4.75, 0) {};
		\node [style=none] (34) at (5.25, 0) {};
		\node [style=none] (35) at (0.5, 2) {};
		\node [style=none] (36) at (0.5, 1) {};
		\node [style=Z dot] (37) at (1.75, 3) {$\alpha_1$};
		\node [style=none] (39) at (8.75, 2.25) {};
		\node [style=Z dot] (40) at (1.75, 2.25) {$\alpha_2$};
		\node [style=Z dot] (41) at (3.5, 4.25) {$k\pi$};
		\node [style=none] (42) at (4.5, -2) {};
		\node [style=none] (43) at (0.5, 3) {};
		\node [style=Z dot rounded rectangle] (45) at (5, -1.25) {$\beta_2 + (k + 1)\pi$};
		\node [style=Z dot] (46) at (7.75, 4.25) {$\varphi$};
		\node [style=none] (47) at (0.5, 2.5) {};
		\node [style=none] (48) at (5.5, -2) {};
		\node [style=Z dot rounded rectangle] (49) at (5, 1) {$\beta_1 + (k + 1)\pi$};
		\node [style=none] (50) at (4.75, 4.25) {};
		\node [style=Z dot rounded rectangle] (51) at (7.75, 3) {$\gamma_1 + k\pi$};
		\node [style=Z dot] (52) at (1.75, 1.5) {$\alpha_3$};
		\node [style=none] (53) at (8.75, 3.25) {};
		\node [style=none] (54) at (8.75, 2.75) {};
		\node [style=none] (55) at (0, 0.75) {$=$};
	\end{pgfonlayer}
	\begin{pgfonlayer}{edgelayer}
		\draw [style=hadamard edge] (0) to (2);
		\draw [style=hadamard edge] (0) to (3);
		\draw [style=hadamard edge] (0) to (4);
		\draw [style=hadamard edge] (0) to (9);
		\draw [style=hadamard edge] (0) to (14);
		\draw (0) to (18.center);
		\draw (1) to (6.center);
		\draw [style=hadamard edge] (1) to (3);
		\draw (1) to (22.center);
		\draw [style=hadamard edge] (2) to (3);
		\draw (2) to (7.center);
		\draw (2) to (19.center);
		\draw [style=hadamard edge] (3) to (9);
		\draw [style=hadamard edge] (3) to (12);
		\draw [style=hadamard edge] (3) to (16);
		\draw (4) to (5.center);
		\draw (4) to (26.center);
		\draw (8.center) to (9);
		\draw (9) to (20.center);
		\draw (12) to (13.center);
		\draw (12) to (24.center);
		\draw (14) to (15.center);
		\draw (14) to (25.center);
		\draw (16) to (17.center);
		\draw (16) to (23.center);
		\draw [style=hadamard edge] (27) to (37);
		\draw [style=hadamard edge] (27) to (40);
		\draw [style=hadamard edge] (27) to (45);
		\draw [style=hadamard edge] (27) to (46);
		\draw [style=hadamard edge] (27) to (49);
		\draw [style=hadamard edge] (27) to (52);
		\draw (28.center) to (32);
		\draw (29.center) to (37);
		\draw (31.center) to (52);
		\draw [in=195, out=15] (32) to (39.center);
		\draw [style=hadamard edge] (32) to (52);
		\draw [style=hadamard edge] (32) to (37);
		\draw [style=hadamard edge] (32) to (40);
		\draw [style=hadamard edge] (32) to (45);
		\draw [style=hadamard edge] (32) to (49);
		\draw (33.center) to (49);
		\draw (34.center) to (49);
		\draw (35.center) to (40);
		\draw (36.center) to (52);
		\draw (37) to (43.center);
		\draw [style=hadamard edge] (37) to (41);
		\draw [style=hadamard edge] (37) to (51);
		\draw [style=hadamard edge] (37) to (45);
		\draw [style=hadamard edge] (37) to (49);
		\draw (40) to (47.center);
		\draw [style=hadamard edge] (40) to (41);
		\draw [style=hadamard edge] (40) to (45);
		\draw [style=hadamard edge] (40) to (49);
		\draw [style=hadamard edge] (40) to (51);
		\draw [style=hadamard edge] (41) to (45);
		\draw [style=hadamard edge] (41) to (52);
		\draw [style=hadamard edge] (41) to (49);
		\draw (42.center) to (45);
		\draw (45) to (48.center);
		\draw [style=hadamard edge] (45) to (52);
		\draw [style=hadamard edge] (45) to (51);
		\draw [style=hadamard edge] (49) to (51);
		\draw [style=hadamard edge] (49) to (52);
		\draw (51) to (54.center);
		\draw (51) to (53.center);
		\draw [style=hadamard edge] (51) to (52);
		\draw [style=hadamard edge] (41) to (50.center);
	\end{pgfonlayer}
\end{tikzpicture}
}
    \caption{Example of the pivot boundary rule}
    \label{fig:pivot_boundary_rule_lhs}
\end{figure}
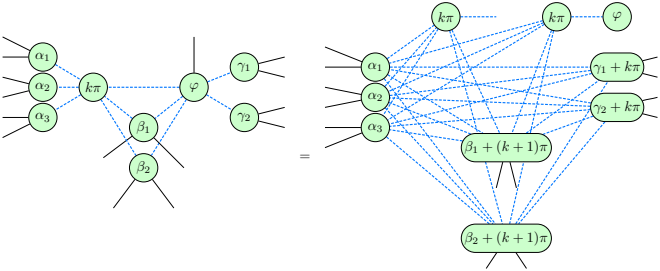

The final rule we have implemented is known as the pivot boundary rule. The difference to the previous two pivot rules is that in this case, we assume that one of the pivot nodes is connected to a simple edge. The rule is demonstrated in Fig.~\ref{fig:pivot_boundary_rule_lhs}.\looseness=-1

\section{Implementation and evaluation}

In this section, we describe the \sys implementation for performing graph rewrites based on the previously described ZX-calculus rewrite rules. The core of the implementation is an extensive set of openCypher queries, which implement the graph rewrite rules. The queries are designed to run on Memgraph, which means they also run on Neo4j with minor modifications. One of the design principles for the queries is to utilize only operations that are supported on both Memgraph and Neo4j.\looseness=-1

\sys implements a method to translate almost any quantum circuit into a ZX-diagram relying on the existing PyZX library. Then, the diagram is imported into the graph database, and certain properties are stored in nodes and edges. These properties are node and edge types, i.e., if the node represents $Z$ or $X$ spider and the phase value. The edges are stored either as simple edges or as Hadamard edges.

The rewrite rules (and queries) can be divided into three categories: independent, weakly independent, and dependent. We can identify substantial performance differences between these categories, which are also demonstrated in the results. We first focus on dependent rewrite rules. We determine that the dependent rules are identity removal in Fig.~\ref{fig:remove_identity}, spider fusion in Fig.~\ref{fig:spider_fusion}, bialgebra rule in Fig.~\ref{fig:bialgebra}, and pivot boundary rule in Fig.~\ref{fig:pivot_boundary_rule_lhs}. A rule is dependent if executing a rewrite based on the rule might affect the rewrite elsewhere in the graph, in the form of deleting nodes or edges. In database terms, this means there is a possibility of conflicting updates of nodes or edges, and without careful design, they will lead to incorrect results.

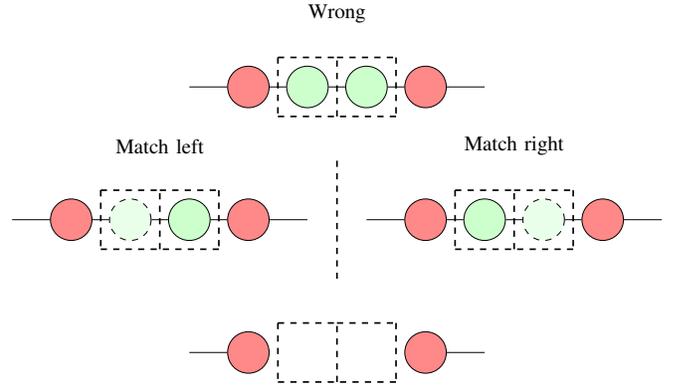
\begin{figure}
    \centering
    \resizebox{\columnwidth}{!}{
\begin{tikzpicture}
	\begin{pgfonlayer}{nodelayer}
		\node [style=Z dot] (0) at (-0.5, 2.25) {};
		\node [style=Z dot] (1) at (0.5, 2.25) {};
		\node [style=X dot] (2) at (-1.5, 2.25) {};
		\node [style=X dot] (3) at (1.5, 2.25) {};
		\node [style=none] (4) at (-2.5, 2.25) {};
		\node [style=none] (5) at (2.5, 2.25) {};
		\node [style=none] (6) at (-1, 2.75) {};
		\node [style=none] (7) at (0, 2.75) {};
		\node [style=none] (8) at (-1, 1.75) {};
		\node [style=none] (9) at (0, 1.75) {};
		\node [style=none] (10) at (1, 2.75) {};
		\node [style=none] (11) at (1, 1.75) {};
		\node [style=Z dot] (13) at (-2.5, 0) {};
		\node [style=X dot] (14) at (-4.5, 0) {};
		\node [style=X dot] (15) at (-1.5, 0) {};
		\node [style=none] (16) at (-5.5, 0) {};
		\node [style=none] (17) at (-0.5, 0) {};
		\node [style=none] (18) at (-4, 0.5) {};
		\node [style=none] (19) at (-3, 0.5) {};
		\node [style=none] (20) at (-4, -0.5) {};
		\node [style=none] (21) at (-3, -0.5) {};
		\node [style=none] (22) at (-2, 0.5) {};
		\node [style=none] (23) at (-2, -0.5) {};
		\node [style=X dot] (25) at (-1.5, -2.25) {};
		\node [style=X dot] (26) at (1.5, -2.25) {};
		\node [style=none] (27) at (-2.5, -2.25) {};
		\node [style=none] (28) at (2.5, -2.25) {};
		\node [style=none] (29) at (-1, -1.75) {};
		\node [style=none] (30) at (0, -1.75) {};
		\node [style=none] (31) at (-1, -2.75) {};
		\node [style=none] (32) at (0, -2.75) {};
		\node [style=none] (33) at (1, -1.75) {};
		\node [style=none] (34) at (1, -2.75) {};
		\node [style=none] (35) at (0, 3.5) {Wrong};
		\node [style=X dot] (71) at (1.5, 0) {};
		\node [style=X dot] (72) at (4.5, 0) {};
		\node [style=none] (73) at (0.5, 0) {};
		\node [style=none] (74) at (5.5, 0) {};
		\node [style=none] (75) at (2, 0.5) {};
		\node [style=none] (76) at (3, 0.5) {};
		\node [style=none] (77) at (2, -0.5) {};
		\node [style=none] (78) at (3, -0.5) {};
		\node [style=none] (79) at (4, 0.5) {};
		\node [style=none] (80) at (4, -0.5) {};
		\node [style=Z dot] (81) at (2.5, 0) {};
		\node [style=none] (82) at (0, 1) {};
		\node [style=none] (83) at (0, -1) {};
		\node [style=Z dot dashed] (84) at (3.5, 0) {};
		\node [style=Z dot dashed] (85) at (-3.5, 0) {};
		\node [style=none] (86) at (-3, 1.25) {Match left};
		\node [style=none] (87) at (3, 1.25) {Match right};
	\end{pgfonlayer}
	\begin{pgfonlayer}{edgelayer}
		\draw (0) to (1);
		\draw (4.center) to (2);
		\draw (2) to (0);
		\draw (1) to (3);
		\draw (3) to (5.center);
		\draw [style=dashed edge] (6.center) to (7.center);
		\draw [style=dashed edge] (6.center) to (8.center);
		\draw [style=dashed edge] (8.center) to (9.center);
		\draw [style=dashed edge] (9.center) to (7.center);
		\draw [style=dashed edge] (7.center) to (10.center);
		\draw [style=dashed edge] (10.center) to (11.center);
		\draw [style=dashed edge] (11.center) to (9.center);
		\draw (16.center) to (14);
		\draw (13) to (15);
		\draw (15) to (17.center);
		\draw [style=dashed edge] (18.center) to (19.center);
		\draw [style=dashed edge] (18.center) to (20.center);
		\draw [style=dashed edge] (20.center) to (21.center);
		\draw [style=dashed edge] (21.center) to (19.center);
		\draw [style=dashed edge] (19.center) to (22.center);
		\draw [style=dashed edge] (22.center) to (23.center);
		\draw [style=dashed edge] (23.center) to (21.center);
		\draw (27.center) to (25);
		\draw (26) to (28.center);
		\draw [style=dashed edge] (29.center) to (30.center);
		\draw [style=dashed edge] (29.center) to (31.center);
		\draw [style=dashed edge] (31.center) to (32.center);
		\draw [style=dashed edge] (32.center) to (30.center);
		\draw [style=dashed edge] (30.center) to (33.center);
		\draw [style=dashed edge] (33.center) to (34.center);
		\draw [style=dashed edge] (34.center) to (32.center);
		\draw (73.center) to (71);
		\draw (72) to (74.center);
		\draw [style=dashed edge] (75.center) to (76.center);
		\draw [style=dashed edge] (75.center) to (77.center);
		\draw [style=dashed edge] (77.center) to (78.center);
		\draw [style=dashed edge] (78.center) to (76.center);
		\draw [style=dashed edge] (76.center) to (79.center);
		\draw [style=dashed edge] (79.center) to (80.center);
		\draw [style=dashed edge] (80.center) to (78.center);
		\draw (71) to (81);
		\draw [style=dashed edge] (82.center) to (83.center);
		\draw (14) to (85);
		\draw (85) to (13);
		\draw (81) to (84);
		\draw (84) to (72);
	\end{pgfonlayer}
\end{tikzpicture}
}
    \caption{One of the simplest possible conflicting patterns in the system, where each identity node is matched separately without ensuring that they are not interacting}
    \label{fig:wrong_match}
\end{figure}

To demonstrate the conflicting, i.e., dependent patterns in their simplest form, consider Fig.~\ref{fig:wrong_match}, where we have four spiders: two red spiders are surrounding two green spiders with phase $0$. By the identity rule in Fig.~\ref{fig:remove_identity}, we can remove both of the green spiders, which should leave two \textit{connected} red spiders. Suppose the query performing this identity-removal rewrite is constructed to match only a single node at a time and remove it. In that case, the system first removes the first green spider, then, in snapshot isolation, the other green spider is removed. Then, the red spiders will never be connected with an edge, since each isolated snapshot assumes that one of the green spiders still exists and connects the wire to that node, while simultaneously removing this node from the other branch of the query. This way, we obtain an incorrect result.

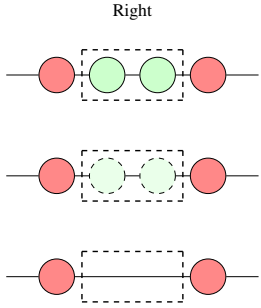
\begin{figure}
    \centering
    \resizebox{0.4\columnwidth}{!}{
\begin{tikzpicture}
	\begin{pgfonlayer}{nodelayer}
		\node [style=Z dot] (0) at (-0.5, 2) {};
		\node [style=Z dot] (1) at (0.5, 2) {};
		\node [style=X dot] (2) at (-1.5, 2) {};
		\node [style=X dot] (3) at (1.5, 2) {};
		\node [style=none] (4) at (-2.5, 2) {};
		\node [style=none] (5) at (2.5, 2) {};
		\node [style=none] (6) at (-1, 2.5) {};
		\node [style=none] (7) at (-1, 1.5) {};
		\node [style=none] (8) at (1, 2.5) {};
		\node [style=none] (9) at (1, 1.5) {};
		\node [style=X dot] (10) at (-1.5, 0) {};
		\node [style=X dot] (11) at (1.5, 0) {};
		\node [style=none] (12) at (-2.5, 0) {};
		\node [style=none] (13) at (2.5, 0) {};
		\node [style=none] (14) at (-1, 0.5) {};
		\node [style=none] (15) at (-1, -0.5) {};
		\node [style=none] (16) at (1, 0.5) {};
		\node [style=none] (17) at (1, -0.5) {};
		\node [style=X dot] (18) at (-1.5, -2) {};
		\node [style=X dot] (19) at (1.5, -2) {};
		\node [style=none] (20) at (-2.5, -2) {};
		\node [style=none] (21) at (2.5, -2) {};
		\node [style=none] (22) at (-1, -1.5) {};
		\node [style=none] (23) at (-1, -2.5) {};
		\node [style=none] (24) at (1, -1.5) {};
		\node [style=none] (25) at (1, -2.5) {};
		\node [style=none] (26) at (0, 3.25) {Right};
		\node [style=Z dot dashed] (27) at (-0.5, 0) {};
		\node [style=Z dot dashed] (28) at (0.5, 0) {};
	\end{pgfonlayer}
	\begin{pgfonlayer}{edgelayer}
		\draw (0) to (1);
		\draw (4.center) to (2);
		\draw (2) to (0);
		\draw (1) to (3);
		\draw (3) to (5.center);
		\draw [style=dashed edge] (6.center) to (7.center);
		\draw [style=dashed edge] (8.center) to (9.center);
		\draw (12.center) to (10);
		\draw (11) to (13.center);
		\draw [style=dashed edge] (14.center) to (15.center);
		\draw [style=dashed edge] (16.center) to (17.center);
		\draw (20.center) to (18);
		\draw (19) to (21.center);
		\draw [style=dashed edge] (22.center) to (23.center);
		\draw [style=dashed edge] (24.center) to (25.center);
		\draw [style=dashed edge] (6.center) to (8.center);
		\draw [style=dashed edge] (7.center) to (9.center);
		\draw [style=dashed edge] (14.center) to (16.center);
		\draw [style=dashed edge] (15.center) to (17.center);
		\draw [style=dashed edge] (22.center) to (24.center);
		\draw [style=dashed edge] (23.center) to (25.center);
		\draw (18) to (19);
		\draw (10) to (27);
		\draw (27) to (28);
		\draw (28) to (11);
	\end{pgfonlayer}
\end{tikzpicture}
}
    \caption{Possible correct matching and rewriting for the interacting patterns}
    \label{fig:right_match}
\end{figure}

\sys tackles the dependent rewrites by first identifying patterns that are not interacting, i.e., they are edge-disjoint, and then rewriting them. The rewrite might introduce more patterns that can be rewritten, so the same query should be executed until no more patterns are found. For example, in the correct matching shown in Fig.~\ref{fig:right_match}, \sys examines larger subgraphs that are guaranteed to remain connected to nodes unaffected by the query rewriting.

Generally, each dependent query starts with a matching pattern that matches either the left-hand or the right-hand side of each rewrite rule. To identify possible conflicting interactions between the neighborhood, the queries first collect candidate matches and then filter out those patterns that are truly disjoint. This filtering phase, which selects independent regions in the graph, is implemented using a reduce function and a structure shown in Fig.~\ref{fig:cypher_example}. The queries in \sys implement a functionality that we would call ''\texttt{(EDGE/NODE) DISJOINT MATCH}'' and return node or edge disjoint patterns. Unfortunately, we are not aware of any graph database that implements this type of query logic, so we had to implement it using the reduce function.

\begin{figure}
    \centering
    \begin{lstlisting}
 WITH reduce(acc = {chosen: [], marked: []}, c IN patterns |
 CASE
   WHEN NONE(nid IN c.neighbors WHERE nid IN acc.marked)
   THEN {chosen: acc.chosen + [c.id], marked: acc.marked + [c.id] + c.neighbors}
   ELSE acc
 END
 ) AS disjoint_patterns
    \end{lstlisting}
    \caption{OpenCypher supported functionality to compute disjoint, non-interacting patterns}
    \label{fig:cypher_example}
\end{figure}

The other categories of queries are constructed for those patterns that are independent or weakly independent. The only independent rule is the phase gadget fusion in Fig.~\ref{fig:gadget_fusion}. A rule (or a query) is independent if its execution does not interact at all with itself during the rewriting process. Interestingly, one can construct a query that identifies arbitrarily large phase gadget patterns and combines them. Due to the nature of the rule, another matching pattern cannot interact with the nodes modified by the query, making this case independent. 

The weakly independent rules are the pivot rule in Fig.~\ref{fig:pivot_rule}, the pivot gadget rule in Fig.~\ref{fig:pivot_gadget_rule}, the pivot boundary rule in Fig.~\ref{fig:pivot_boundary_rule_lhs}, and the local complementation rule in Fig.~\ref{fig:local_complementation_rule}. We call a rule (or a query) weakly independent if the rewrite may modify properties of a node or an edge that is part of another pattern to which the rewrite applies. Compared to the dependent rules, these rules do not interact in a way that would cause a node or an edge, which is to be deleted, to be shared between two different matching patterns. Moreover, we practically considered these rules independent, since there are only exceptional corner cases in which conflicts can occur.

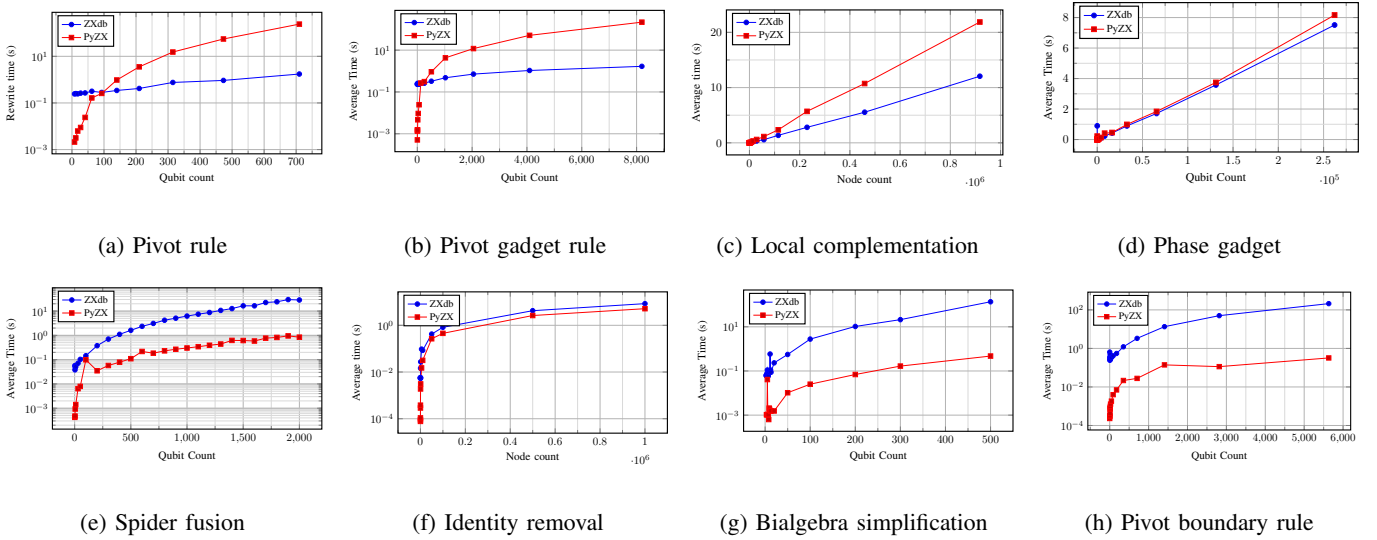
\begin{figure*}[ht]
    \centering
    
    \begin{subfigure}{0.24\textwidth}
        \centering
        \resizebox{\columnwidth}{!}{
\begin{tikzpicture}
  \begin{axis}[
    width=10cm,
    height=6cm,
    xlabel={Qubit count},
    ylabel={Rewrite time (s)},
    grid=both,
    grid style={line width=.1pt, draw=gray!30},
    major grid style={line width=.2pt,draw=gray!60},
    minor tick num=1,
    legend style={at={(0.02,0.98)},anchor=north west, font=\small},
    every axis plot/.append style={thick},
    ymode=log,
  ]
    \addplot+[mark=*, color=blue] coordinates {(8, 0.2442624568939209)
(12, 0.25238823890686035)
(18, 0.24572420120239258)
(27, 0.2635471820831299)
(41, 0.2689087390899658)
(62, 0.31762146949768066)
(93, 0.28235578536987305)
(140, 0.34362339973449707)
(210, 0.418398380279541)
(315, 0.7577545642852783)
(473, 0.9255285263061523)
(710, 1.7389552593231201)};
    \addlegendentry{ZXdb}
    \addplot+[mark=square*, color=red] coordinates {(8, 0.002107858657836914)
(12, 0.0031681060791015625)
(18, 0.00624847412109375)
(27, 0.00887441635131836)
(41, 0.023962020874023438)
(62, 0.1641843318939209)
(93, 0.2568700313568115)
(140, 0.9652762413024902)
(210, 3.535442590713501)
(315, 15.270036935806274)
(473, 55.39441704750061)
(710, 240.4711685180664)};
    \addlegendentry{PyZX}
  \end{axis}
\end{tikzpicture}
}
        \caption{Pivot rule}
        \label{fig:pivot_results}
    \end{subfigure}\hfill
    \begin{subfigure}{0.24\textwidth}
        \centering
        \resizebox{\columnwidth}{!}{
\begin{tikzpicture}
  \begin{axis}[
    width=10cm,
    height=6cm,
    xlabel={Qubit Count},
    ylabel={Average Time (s)},
    grid=both,
    grid style={line width=.1pt, draw=gray!30},
    major grid style={line width=.2pt,draw=gray!60},
    minor tick num=1,
    legend style={at={(0.02,0.98)},anchor=north west, font=\small},
    every axis plot/.append style={thick},
    ymode=log,
  ]
    \addplot+[mark=*, color=blue] coordinates {(2, 0.24240994453430176)
(3, 0.23552393913269043)
(5, 0.23113401730855307)
(9, 0.2574036121368408)
(17, 0.23092103004455566)
(33, 0.23255157470703125)
(65, 0.23687076568603516)
(129, 0.25920557975769043)
(257, 0.262423038482666)
(513, 0.3262796401977539)
(1025, 0.47759413719177246)
(2049, 0.7143020629882812)
(4097, 1.063962697982788)
(8193, 1.677936315536499)};
    \addlegendentry{ZXdb}
    \addplot+[mark=square*, color=red] coordinates {(2, 0.0005090236663818359)
(3, 0.0013418197631835938)
(5, 0.0015526612599690754)
(9, 0.0015134811401367188)
(17, 0.0046596527099609375)
(33, 0.009344339370727539)
(65, 0.024440526962280273)
(129, 0.2666182518005371)
(257, 0.31214094161987305)
(513, 0.9239795207977295)
(1025, 4.2828528881073)
(2049, 11.861563682556152)
(4097, 50.954328298568726)
(8193, 217.63900423049927)};
    \addlegendentry{PyZX}
  \end{axis}
\end{tikzpicture}
}
        \caption{Pivot gadget rule}
        \label{fig:pivot_gadget_results}
    \end{subfigure}\hfill
        \begin{subfigure}{0.24\textwidth}
        \centering
        \resizebox{\columnwidth}{!}{
\begin{tikzpicture}
  \begin{axis}[
    width=10cm,
    height=6cm,
    xlabel={Node count},
    ylabel={Average time (s)},
    grid=both,
    grid style={line width=.1pt, draw=gray!30},
    major grid style={line width=.2pt,draw=gray!60},
    minor tick num=1,
    legend style={at={(0.02,0.98)},anchor=north west, font=\small},
    every axis plot/.append style={thick},
  ]
    \addplot+[mark=*, color=blue] coordinates {(20, 0.007170700002461672)
(34, 0.006819600006565452)
(62, 0.005570999986957759)
(118, 0.0072465999983251095)
(230, 0.008328600000822917)
(454, 0.01055200002156198)
(902, 0.01567270001396537)
(1798, 0.02454849999048747)
(3590, 0.05175959999905899)
(7174, 0.1038966000196524)
(14342, 0.18761709998943843)
(28678, 0.3444782999868039)
(57350, 0.5739059999759775)
(114694, 1.3567331000231206)
(229382, 2.8076349000039045)
(458758, 5.543700799986254)
(917510, 12.056602199998451)};
    \addlegendentry{ZXdb}
    \addplot+[mark=square*, color=red] coordinates {(20, 0.0004250000056345016)
(34, 0.0010100000072270632)
(62, 0.0011587999761104584)
(118, 0.0022431000252254307)
(230, 0.0042181000171694905)
(454, 0.007845999993151054)
(902, 0.016344200004823506)
(1798, 0.03145030001178384)
(3590, 0.06361840001773089)
(7174, 0.12680170001112856)
(14342, 0.3327264999970794)
(28678, 0.6000470999861136)
(57350, 1.1306818000157364)
(114694, 2.3570267999893986)
(229382, 5.698524599982193)
(458758, 10.733551099983742)
(917510, 21.869801499997266)};
    \addlegendentry{PyZX}
  \end{axis}
\end{tikzpicture}
}
        \caption{Local complementation}
        \label{fig:local_complementation_results}
    \end{subfigure}
    \hfill
        \begin{subfigure}{0.24\textwidth}
        \centering
        \resizebox{\columnwidth}{!}{
\begin{tikzpicture}
  \begin{axis}[
    width=10cm,
    height=6cm,
    xlabel={Qubit Count},
    ylabel={Average Time (s)},
    grid=both,
    grid style={line width=.1pt, draw=gray!30},
    major grid style={line width=.2pt,draw=gray!60},
    minor tick num=1,
    legend style={at={(0.02,0.98)},anchor=north west, font=\small},
    every axis plot/.append style={thick},
  ]
    \addplot+[mark=*, color=blue] coordinates {(1, 0.8975282666644835)
(2, 0.0068137999915052205)
(4, 0.005614099995000288)
(8, 0.005756599974120036)
(16, 0.00510109998867847)
(32, 0.00565679999999702)
(64, 0.006853999977465719)
(128, 0.007971450002514757)
(256, 0.010243899974739179)
(512, 0.018778500001644716)
(1024, 0.026055800000904128)
(2048, 0.05027370000607334)
(4096, 0.09281000000191852)
(8192, 0.22910769999725744)
(16384, 0.4266766499931691)
(32768, 0.8919947499962291)
(65536, 1.709502199984854)
(131072, 3.5778464000031818)
(262144, 7.511175199993886)};
    \addlegendentry{ZXdb}
    \addplot+[mark=square*, color=red] coordinates {(1, 0.2206920999978181)
(2, 0.00018569998792372644)
(4, 0.00020000000949949026)
(8, 0.0003507000219542533)
(16, 0.0005163000023458153)
(32, 0.0008203999896068126)
(64, 0.001382599992211908)
(128, 0.003749550014617853)
(256, 0.009140699985437095)
(512, 0.020191599993268028)
(1024, 0.02531789999920875)
(2048, 0.05246459998306818)
(4096, 0.1294630999909714)
(8192, 0.412389200006146)
(16384, 0.4578470999986166)
(32768, 0.9891885999822989)
(65536, 1.8399499999941327)
(131072, 3.7445380999997724)
(262144, 8.164061599993147)};
    \addlegendentry{PyZX}
  \end{axis}
\end{tikzpicture}
}
        \caption{Phase gadget}
        \label{fig:phase_gadget_results}
    \end{subfigure}

    \vspace{1em}
    \begin{subfigure}{0.24\textwidth}
        \centering
        \resizebox{\columnwidth}{!}{
\begin{tikzpicture}
  \begin{axis}[
    width=10cm,
    height=6cm,
    xlabel={Qubit Count},
    ylabel={Average Time (s)},
    grid=both,
    grid style={line width=.1pt, draw=gray!30},
    major grid style={line width=.2pt,draw=gray!60},
    minor tick num=1,
    legend style={at={(0.02,0.98)},anchor=north west, font=\small},
    every axis plot/.append style={thick},
    ymode=log,
  ]
    \addplot+[mark=*, color=blue] coordinates {(2, 0.05679330002749339)
(3, 0.038011400029063225)
(5, 0.048616699990816414)
(10, 0.051827099989168346)
(30, 0.07091050001326948)
(50, 0.10262190000503324)
(100, 0.1452030999935232)
(200, 0.3758769999840297)
(300, 0.7021667999797501)
(400, 1.103978700004518)
(500, 1.615799600025639)
(600, 2.351525900012348)
(700, 3.1033001000178047)
(800, 4.203019100008532)
(900, 5.099897800013423)
(1000, 6.2312293000286445)
(1100, 7.450460399966687)
(1200, 8.745294799969997)
(1300, 10.657877999998163)
(1400, 12.67952459998196)
(1500, 16.462696500006132)
(1600, 16.43752720003249)
(1700, 22.557530599995516)
(1800, 24.11913020000793)
(1900, 29.72031830000924)
(2000, 28.621278400009032)};
    \addlegendentry{ZXdb}
    \addplot+[mark=square*, color=red] coordinates {(2, 0.0004216000088490546)
(3, 0.00047670002095401287)
(5, 0.0009285000269301236)
(10, 0.0014121999847702682)
(30, 0.006392200011759996)
(50, 0.007922699995106086)
(100, 0.0986393999774009)
(200, 0.0348431000020355)
(300, 0.05741100001614541)
(400, 0.07865229999879375)
(500, 0.10955499997362494)
(600, 0.2173274999950081)
(700, 0.18474580004112795)
(800, 0.23037800000747666)
(900, 0.2680875000078231)
(1000, 0.30207710003014654)
(1100, 0.3416459999862127)
(1200, 0.3912022999720648)
(1300, 0.4396245999960229)
(1400, 0.6203171999659389)
(1500, 0.6121343999984674)
(1600, 0.585294300050009)
(1700, 0.7678782999864779)
(1800, 0.8224723000312224)
(1900, 0.9513904000050388)
(2000, 0.8432996000046842)};
    \addlegendentry{PyZX}
  \end{axis}
\end{tikzpicture}
}
        \caption{Spider fusion}
        \label{fig:spider_fusion_results}
    \end{subfigure}\hfill
    \begin{subfigure}{0.24\textwidth}
        \centering
        \resizebox{\columnwidth}{!}{
\begin{tikzpicture}
  \begin{axis}[
    width=10cm,
    height=6cm,
    xlabel={Node count},
    ylabel={Average time (s)},
    grid=both,
    grid style={line width=.1pt, draw=gray!30},
    major grid style={line width=.2pt,draw=gray!60},
    minor tick num=1,
    legend style={at={(0.02,0.98)},anchor=north west, font=\small},
    every axis plot/.append style={thick},
    ymode=log,
  ]
    \addplot+[mark=*, color=blue] coordinates {(4, 0.005678099987562746)
(7, 0.005732599995099008)
(12, 0.0052630999707616866)
(52, 0.005781400017440319)
(102, 0.005565400002524257)
(502, 0.014626300020609051)
(1002, 0.027238700015004724)
(5002, 0.09543410001788288)
(10002, 0.0845644999644719)
(50002, 0.4210636000498198)
(100002, 0.8217991999699734)
(500002, 4.223216000013053)
(1000002, 8.460597199969925)};
    \addlegendentry{ZXdb}
    \addplot+[mark=square*, color=red] coordinates {(4, 7.700000423938036e-05)
(7, 9.01000457815826e-05)
(12, 0.00010839995229616761)
(52, 0.0002883999841287732)
(102, 0.00038079998921602964)
(502, 0.0018850000342354178)
(1002, 0.0030920999706722796)
(5002, 0.014955799968447536)
(10002, 0.03156410000519827)
(50002, 0.26508450001711026)
(100002, 0.45292230002814904)
(500002, 2.6214114000322297)
(1000002, 5.142916500044521)};
    \addlegendentry{PyZX}
  \end{axis}
\end{tikzpicture}
}
        \caption{Identity removal}
        \label{fig:identity_removal_results}
    \end{subfigure}\hfill
    \begin{subfigure}{0.24\textwidth}
        \centering
        \resizebox{\columnwidth}{!}{
\begin{tikzpicture}
  \begin{axis}[
    width=10cm,
    height=6cm,
    xlabel={Qubit Count},
    ylabel={Average Time (s)},
    grid=both,
    grid style={line width=.1pt, draw=gray!30},
    major grid style={line width=.2pt,draw=gray!60},
    minor tick num=1,
    legend style={at={(0.02,0.98)},anchor=north west, font=\small},
    every axis plot/.append style={thick},
    ymode=log,
  ]
    \addplot+[mark=*, color=blue] coordinates {(2, 0.06359386444091797)
(3, 0.059434096018473305)
(4, 0.07566225528717041)
(5, 0.11014747619628906)
(6, 0.06879878044128418)
(7, 0.059044599533081055)
(8, 0.10355941454569499)
(9, 0.08659696578979492)
(10, 0.10336653391520183)
(11, 0.585033655166626)
(12, 0.11058712005615234)
(13, 0.08728456497192383)
(20, 0.23302841186523438)
(50, 0.5594429969787598)
(100, 2.7809383869171143)
(200, 10.38192629814148)
(300, 21.131150245666504)
(500, 136.2111690044403)};
    \addlegendentry{ZXdb}
    \addplot+[mark=square*, color=red] coordinates {(2, 0.0)
(3, 0.0010494391123453777)
(4, 0.0010346174240112305)
(5, 0.04086152712504069)
(6, 0.001085519790649414)
(7, 0.0010478496551513672)
(8, 0.0006329218546549479)
(9, 0.0020775794982910156)
(10, 0.0018288294474283855)
(11, 0.0)
(12, 0.001547098159790039)
(13, 0.0015690326690673828)
(20, 0.0015437602996826172)
(50, 0.01033329963684082)
(100, 0.025181055068969727)
(200, 0.06919240951538086)
(300, 0.16641998291015625)
(500, 0.4740579128265381)};
    \addlegendentry{PyZX}
  \end{axis}
\end{tikzpicture}
}
        \caption{Bialgebra simplification}
        \label{fig:bialgebra_simp_results}
    \end{subfigure}\hfill
    \begin{subfigure}{0.24\textwidth}
        \centering
        \resizebox{\columnwidth}{!}{
\begin{tikzpicture}
  \begin{axis}[
    width=10cm,
    height=6cm,
    xlabel={Qubit Count},
    ylabel={Average Time (s)},
    grid=both,
    grid style={line width=.1pt, draw=gray!30},
    major grid style={line width=.2pt,draw=gray!60},
    minor tick num=1,
    legend style={at={(0.02,0.98)},anchor=north west, font=\small},
    every axis plot/.append style={thick},
    ymode=log,
  ]
    \addplot+[mark=*, color=blue] coordinates {(1, 0.647072149993619)
(3, 0.24239600001601502)
(4, 0.3575044000026537)
(8, 0.28684460000658873)
(11, 0.4874322999967262)
(12, 0.29172325000399724)
(14, 0.3400932000222383)
(15, 0.25986859999829903)
(22, 0.32100315000570845)
(44, 0.331949999963399)
(88, 0.4127151499997126)
(176, 0.547537500038743)
(352, 1.2134914000052959)
(704, 3.3238069000071846)
(1408, 13.458722500014119)
(2816, 50.12674519995926)
(5632, 211.32521549996454)};
    \addlegendentry{ZXdb}
    \addplot+[mark=square*, color=red] coordinates {(1, 0.00023099999816622585)
(3, 0.00029269998776726425)
(4, 0.0003205499961040914)
(8, 0.00042055001540575176)
(11, 0.0007757000275887549)
(12, 0.0007266500033438206)
(14, 0.000830300006782636)
(15, 0.000987899984465912)
(22, 0.0012909499928355217)
(44, 0.001840100041590631)
(88, 0.003973099999711849)
(176, 0.0070235999883152544)
(352, 0.021626200003083795)
(704, 0.027838599984534085)
(1408, 0.13968890003161505)
(2816, 0.11344280000776052)
(5632, 0.3205479999887757)};
    \addlegendentry{PyZX}
  \end{axis}
\end{tikzpicture}
}
        \caption{Pivot boundary rule}
        \label{fig:pivot_boundary_results}
    \end{subfigure}

    \vspace{0.5em}
    \caption{Performance results for the implemented rewrite rules. Each subfigure shows the runtime behavior for one specific rule.\looseness=-1}
    \label{fig:all_performance_results}
\end{figure*}

\subsection{Evaluation}

The evaluation is divided into two parts: ensuring the correctness of the implemented queries and comparing system performance with the PyZX library. Correctness is compared against the rewritten graphs from PyZX and the original circuits. In cases where the sequence of rule applications might lead to different graphs implementing the same operation, we compare the graphs as mappings defined by their underlying tensor networks~\cite{biamonte2017tensornetworksnutshell}. The core idea is that the rewrite rules do not change the underlying linear mapping that the graph or the circuit implements. 

\subsubsection{Corretness}
The correctness of the rewrite rules implemented as queries is ensured at multiple levels, depending on the graph, i.e, quantum circuit size. From the most costly to the least expensive method, the correctness tests cover tensor-level comparison, graph isomorphism, and degree sequences.

\paragraph{Tensor-level comparison}
Every ZX-diagram and quantum circuit can be represented by a tensor network, which corresponds to a special high-dimensional tensor. \sys converts ZX-diagrams with an even number of inputs and outputs into tensor networks, which are contracted with Quimb software~\cite{Gray2018}. This is a substantially more scalable package compared to the tensor module that PyZX currently implements, and improves the ZX-diagram simulation in this regard. The output of this process is an operator, which is compared with the operator-level equivalence using Qiskit. We also take into account that the rewrite rules and other transformations might cause qubit rotations, so the system automatically checks operators with respect to these qubit permutations. We have also implemented mechanisms that compare the corresponding tensors based on their spectral characteristics and traces of operator powers.

\paragraph{Graph isomorphism}
Graph isomorphism is a robust and relatively scalable method to define if the \sys rewritten graphs match the PyZX written graphs. If the graphs are isomorphic, then they necessarily represent the same mapping, i.e., quantum circuit or tensor network. In the implementation, computing isomorphism also involves considering node types ($Z$- and $X$-spiders), edge types (simple wires and Hadamard wires), as well as the phase values. If the graph structures and properties match exactly, the outputs are isomorphic, and thus the transformations are correct. Graph isomorphism can be solved for larger graphs than the exact tensor network simulation scales.

\paragraph{Degree sequences}
Degree sequences can be considered as a heuristic to determine if the \sys rewritten graphs match the PyZX written graphs. If the degree sequences for the graphs are the same, the transformation has likely been the same for a particular set of rules. Since computing the degree sequence requires only $\mathcal{O}(n)$ steps in terms of the nodes, this is a scalable heuristic between graphs for which we cannot compute the previous two metrics.

\subsubsection{Performance}
The performance statistics are collected from a work station equipped with AMD Ryzen 9 7945HX CPU and 32 GB RAM. Both PyZX and MemGraph are capable of utilizing only a single CPU, and MemGraph is running in a Docker environment. The main performance results are collected in Fig.~\ref{fig:all_performance_results}. The first row contains those rewrite rules that are independent or weakly independent, and the second row contains those rules that are dependent. The results show a performance difference between these two categories. By profiling the queries in Memgraph, it becomes clear that the main reason for the performance difference arises from the utilization of the reduce function and the pattern shown in Fig.~\ref{fig:cypher_example}. This pattern can take $80-95\%$ of the execution time. Thus, the graph databases would become more feasible for these types of rewriting tasks if they supported edge and node disjoint pattern matching. Nevertheless, for a substantial subset of the rules, graph databases appear as a fast and scalable alternative to perform ZX rewrite rules.

\section{Discussion}

In this work, we focus on a well-defined subset of operations, namely ZX-calculus rewrite rules, which many modern quantum compilation systems utilize to reason, simplify, and optimize quantum circuits. Considering the list of possible operations we outlined in the background section, quantum circuit compilation appears to be a field that will likely benefit and scale by relying on database management systems at multiple stages. In addition to expressing quantum circuits as graphs using ZX-diagrams, another standard formalism is to write a circuit as a Directed Acyclic Graph (DAG), which describes causal relationships between the quantum logical operations in the circuit~\cite{Meijer_van_de_Griend_2025}. This formalism is commonly used in Qiskit~\cite{qiskit-transpiler-doc}, Pennylane~\cite{pennylane-transforms-doc}, and Tket~\cite{tket-comp-doc} among other quantum software development packages. In future work, we will also include support for DAGs in \sys, which are particularly useful for querying circuit statistics, decomposing gates, and rewriting quantum circuits using hardware-native gates. On the other hand, topics such as resynthesis, routing, and optimization are more linear algebraic than graph theoretic, and thus transferring those workloads to databases might be more complex. 

Databases may also naturally support multiple other quantum computing pipelines~\cite{10.1145/3736393.3736694}. Although the suitability of databases as quantum simulators is debatable, since relational databases do not naturally support operations such as singular value decomposition, which is fundamental to quantum simulation, numerous other promising opportunities remain. Quantum computers will generally produce vast amounts of measurement and calibration data, which will be used, for example, in error mitigation algorithms~\cite{uotila2025perspectives}. Fault-tolerant quantum computing and compilation for the fault-tolerant devices~\cite{Paler_Polian_Nemoto_Devitt_2017} might also create new data processing challenges. 

The implementation of \sys reveals several interesting characteristics of the capabilities of the current graph databases. Graph rewrites are a special type of workload that graph databases do not yet smoothly support. One might assume that graph database updates become a bottleneck, but we found that implementing sequential structures, such as the reduce operation in Fig.~\ref{fig:cypher_example}, was the most expensive. This was used to ensure the disjointness of the patterns that will be rewritten. Secondly, the current workloads, such as non-interacting disjoint updates, are theoretically easy to parallelize, but graph databases currently do not support parallelism extensively. While Neo4j has a parallel runtime~\cite{PryceAaklundhAverbuch2023}, Memgraph is still in the process of developing it. For database developers not working with quantum computing, \sys provides a write-heavy graph database benchmark platform that scales naturally to billions of nodes and implements a set of complex openCypher queries.

\section{Conclusion}

This work introduces \sys, which implements the ZX-calculus rewrite rules on graph databases. We have developed and demonstrated the feasibility of utilizing graph databases and graph queries in circuit compilation by rewriting circuits using ZX-calculus and openCypher queries on the Memgraph database. Depending on whether the rewrites are independent or dependent, we can see that the graph database implementation outperforms the baseline framework in graph rewriting. 

Quantum computing technologies will be producing increasing amounts of complex data in the future. Storing and utilizing this data will likely help tackle various challenges in quantum technological pipelines, such as compilation, error mitigation, and error correction. Efficient storage and querying of this data create new challenges in data management. In this work, we tackled one of these fundamental challenges by introducing a system that efficiently queries and updates quantum circuits, providing a feasible demonstration for integrating data management into the quantum computing workflow.

\section{Acknowledgments}
This work is funded by Research Council of Finland (grant number 362729), Business Finland (grant number 169/31/2024), and the \emph{Finnish Quantum Flagship Exploratory Project} to PI Bo Zhao. 

\balance
\bibliographystyle{IEEEtran}
\bibliography{refs.bib}

\end{document}